\documentclass[12pt,preprint]{aastex}

\usepackage{epsfig}
\usepackage{amssymb}
\usepackage{makeidx}
\usepackage{ifthen}
\usepackage{verbatim}
\usepackage{natbib}

\begin{document}

\title{Post Main Sequence Orbital Circularization of Binary Stars in the Large and Small Magellanic Clouds.}

\author{Lorenzo Faccioli \altaffilmark{1}}
\affil{Department of Physics and Astronomy, University of Pennsylvania, Philadelphia, PA 19104, USA
and Lawrence Berkeley National Laboratory, Berkeley, CA 94720, USA}

\author{Charles Alcock \altaffilmark{2}} 
\affil{Harvard-Smithsonian Center for Astrophysics, Cambridge, MA 02138, USA}

\and

\author{Kem Cook \altaffilmark{3}}
\affil{Lawrence Livermore National Laboratory, Livermore, CA 94550, USA}

\altaffiltext{1}{\textsl{LFaccioli@lbl.gov}}
\altaffiltext{2}{\textsl{calcock@cfa.harvard.edu}}
\altaffiltext{3}{\textsl{kcook@igpp.ucllnl.org}}
\begin{abstract}
We present results from a study of the orbits of eclipsing binary stars (EBs) in the Magellanic Clouds.
The samples comprise $4510$ EBs found in the Large Magellanic Cloud (LMC) by the MACHO project, $2474$ LMC EBs found by the OGLE-II project (of which $1182$ are also in the MACHO sample), $1380$ in the Small Magellanic Cloud (SMC) found by the MACHO project,
and $1317$ SMC EBs found by the OGLE-II project (of which $677$ are also in the MACHO sample); we also consider the EROS sample of $79$ EBs in the bar of the LMC.
Statistics of the phase differences between primary and secondary minima allow 
us to infer the statistics of orbital eccentricities within these samples.
We confirm the well-known absence of eccentric orbit in close binary stars.
We also find evidence for rapid circularization in longer period systems when one member evolves beyond the main sequence, as also found by previous studies.
\end{abstract}
\section{Introduction}
The two components of a binary star system raise tides in each other.
These tides will, in general, cause dissipation in the stars, leading to the phenomenon of \emph{tidal lag}.
These lags exert torques which lead to exchange of angular momentum between the orbit and the spins of the stars.
The lowest energy configuration, for a given total angular momentum, is a circular orbit with both spins aligned and synchronous with the orbit.
All other configurations evolve towards the synchronous, circular state.
Whether or not this is achieved is determined by the rate of dissipation in the tides.
\par
The study of tides raised in celestial bodies by mutual attraction dates back at least to \citet{darwin79}.
Darwin considered the tides raised on the Earth, modeled as a homogeneous and deformable \emph{viscous} body, by a point mass Moon in the weak friction limit; hence his model is also known as \emph{weak friction model}.
The model assumes that viscous dissipation causes a delay in the onset of the tide by a constant amount $\tau$ so at time $t$ the shape of the Earth is the one that it would have been at time $t-\tau$ in the absence of dissipation.
The axis of the tidal bulge is not therefore aligned with the line of the centers of the two bodies but lags it by a constant amount, resulting in a torque that tends to align the two bodies.
Modern discussions of the weak friction model, with emphasis on close binary stars rather than planet-satellite systems, are given by \citet{alexander73} and \citet{hut81}.
\par
\citet{zahn66a,zahn66b,zahn66c} proposed that the coupling of the tidal 
flow with turbulent flows in the envelope of a late type star is chiefly responsible
for orbital circularization and synchronization in these stars; evidence of circularization due to tidal interactions in late type giants is found by \citet{lucy71}.
The theory has been more recently revised by \citet{zahn89a} and compared against observations by \citet{zahn89b} who found that for late type stars most of the circularization occurs during the Hayashi phase of pre main sequence evolution, as previously suggested by \citet{mayor84}.
\par
The circularization models cited above consider only the \emph{equilibrium tide} which arises if the star is at all times in hydrostatic equilibrium.
If, however, the orbit is not circular or the rotation is not synchronous the star is subjected to a time-varying gravitational potential which excites oscillations, giving rise to a \emph{dynamical tide} superimposed on the equilibrium tide.
Forced oscillations in binary stars were first considered by \citet{cowling41} and then by \citet{zahn70}; \citet{zahn75} considered the damping of the dynamical tide by radiative dissipation in the radiative envelope as a possible circularization mechanism for early type stars; \citet{giuricin84} studied a sample of $\sim 200$ early type binary stars and showed the theory to be compatible with the data presented.
These mechanisms are all described in \citet{zahn77}, where time scales of 
circularization and synchronization are derived.
\par
A circularization mechanism in which energy dissipation is due to large scale hydro-dynamical flows resulting from the deformation of the star (meridional circulation) has been proposed by Tassoul \citep{tassoul87,tassoul95}.
This model has been applied by \citet{claret95} to a homogeneous sample of 45 eclipsing binary stars (EBs) with accurate parameter determination from \citet{andersen91}: they find a satisfactory agreement with the observations and also find that circularization is still taking place during the main sequence for early-type EBs, a finding consistent with the results of this paper.
\citet{claret97} then apply both the turbulent dissipation mechanism and the radiative damping mechanism to the same \citep{andersen91} data set.
They find that within uncertainties, these formalisms seem able to explain the observed eccentricity distribution, although with a few exceptions.
\par
Since tidal forces decrease and the period increases with separation, in a sample of coeval binary stars such as those in star clusters, binaries with a longer period should have orbits with a range of eccentricities since tidal forces have not circularized them yet, whereas binaries with a shorter period should all have circular orbits since tidal forces would have been more effective in circularizing them.
Therefore clusters should show a transition between binaries with shorter period and circular orbits and binaries with longer periods and eccentric orbits; the values of this transition period varies with the age of the cluster.
Binary stars in cluster are widely studied: for example as a part of the WIYN Open Cluster Study (WOCS) \footnote{\url{http://www.astro.ufl.edu/\~{}ata/wocs/}}, a systematic 
search of EBs in Milky Way open clusters for testing orbital circularization theories is currently under way \citep{mathieu04,meibom05}.
\citet{latham02} present spectroscopic orbital solutions for $171$ single-lined stars from their catalog \citep{carney94} of $1464$ stars selected for high proper motion and find that for the metal poor, high velocity halo binary stars in their sample the transition from circular to eccentric orbits occurs at $\sim 20~\mathrm{days}$.
\par
Binary systems in which the component stars eclipse each other are a powerful tool for the study of circularization theories since in this case both \emph{stellar} parameters like radius, mass, and temperature, and \emph{orbital} parameters like eccentricity, angle of inclination, and longitude of periastron, can be determined to the accuracy necessary to test different theories.
To determine these parameters fully one needs to supplement the photometric data with high quality spectra, which can be used to determine orbital velocities.
\par
Eclipsing binary stars have been found in large numbers by astronomical surveys studying gravitational microlensing: MACHO\footnote{\url{http://www.macho.mcmaster.ca/}}, OGLE\footnote{\url{http://sirius.astrouw.edu.pl/\~{}ogle/}} and EROS\footnote{\url{http://eros.in2p3.fr/}}.
In particular a sample of $4634$ EBs in the Large Magellanic Cloud (LMC) and of $1509$ EBs in the Small Magellanic Cloud \citep{faccioli07} have just been published by the MACHO collaboration.
Previously a catalogue of $2580$ EBs in the LMC \citep{wyr03} and of $1351$ EBs in the SMC \citep{wyr04} were published by the OGLE collaboration, the EBs were selected from their catalogue of variable stars in the Magellanic Clouds \citep{zebrun01a} compiled from observations taken during the second part of the project \citep[OGLE-II:][]{udalski97}.
A catalogue of 611 EBs in the LMC \citep{lacy97}, and of $79$ EBs in the bar of the LMC \citep{gri95} were also published by the MACHO and EROS collaboration respectively.
This earlier MACHO sample is contained in the new one.
\par
We aim to take advantage of the sheer size of our samples to infer statistically valid conclusions on orbital circularization, even though our data do not have high photometric precision and there are no spectroscopic data. 
We exploit a simple idea: the phase difference between the primary and secondary minima in an EB light curve, $\phi_1-\phi_2$, may be effectively estimated for the light curves in 
our samples.
This object is related to the orbital elements by the following equation, which sets a \emph{lower bound} to the eccentricity \citep{milone98}:
\begin{equation}\label{eq:ecc}
|\phi_1-\phi_2|-\frac{1}{2}=
\frac{1}{\pi}e\cos\omega\left(1+\frac{1}{\sin^2i}\right)
\end{equation}
where $\phi_1-\phi_2$ is the phase difference between primary and secondary minima, $e$ is the eccentricity, $\omega$ is the argument of periastron and $i$ is the orbital inclination.
For eclipsing systems, especially the wider systems where non zero eccentricities are found, we know that $\sin i\sim 1$.
The important degeneracy is between $\omega$ and $e$.
It is possible \emph{in principle} to model high signal-to-noise 
photometric data to break this degeneracy \citep{wilson01,wilson02,devor05}
but we do not attempt this here.
We will show that $\phi_1-\phi_2$ alone may be used for our purpose.
Although Eq. \ref{eq:ecc} shows that for an eccentric orbit $\phi_1-\phi_2$
can be either $<0.5$ or $>0.5$, depending on $\cos\omega$, we will always 
adopt the convention $\phi_1-\phi_2>0.5$ since only the deviation of $\phi_1-\phi_2$ from $0.5$ is relevant for the detection of eccentricity.
\par
The paper is organized as follows: Section \ref{sec:obs} describes both the MACHO and the OGLE-II samples for both Clouds; Section \ref{sec:fit} describes the fits to the MACHO data and its validity (for the OGLE-II data the relevant information is provided by the authors); Section \ref{sec:results} reports our results for both Clouds both from the MACHO and the OGLE-II samples; Section \ref{sec:sig} discusses the significance of our results and EROS results are briefly considered in Subsection \ref{subsec:eros}. 
Finally Section \ref{sec:conclusion} states our conclusions. 
\par
The data presented in this paper can be accessed on line at the Astrophysical Journal
website\footnote{\url{http://www.journals.uchicago.edu/ApJ/}}
and are mirrored at the Harvard University Initiative in Innovative Computing (IIC)
/Time Series Center\footnote{\url{http://timemachine.iic.harvard.edu/faccioli/CircularizationTables/}}.
\par
From now on we will always use the term \emph{unfolded light curve} to indicate a set of time ordered observations and will reserve the term \emph{light curve} to indicate a set of time ordered observations \emph{folded} around a period, omitting for brevity the adjectives ``folded'' or ``phased''.
\section{Data sets}
\label{sec:obs}
\subsection{MACHO data}
The MACHO Project was an astronomical survey whose primary aim was to detect gravitational microlensing events of background stars by compact objects in the halo of the Milky Way.
The background stars were located in the LMC, SMC and the bulge of the Milky Way; more details on the detection of microlensing events can be found in \citet{alcock00} and references therein.
\par
Observations were carried out from July 1992 to December 1999 with the dedicated $1.27\mathrm{m}$ telescope of Mount Stromlo, Australia, using a $2\times 2$ mosaic of $2048\times 2048$ CCD in two band passes simultaneously.
These are called MACHO ``blue", hereafter indicated with $V_{\mathrm{MACHO}}$, with a bandpass of $440-590\mathrm{nm}$ and MACHO ``red", hereafter indicated with $R_{\mathrm{MACHO}}$, with a bandpass of $590-780\mathrm{nm}$; transformations to standard Johnson $V$ and Cousins $R$ bands are described in detail in \citet{alcock99}; the magnitudes quoted in this paper have been obtained by using the following transformation for the LMC:
\begin{eqnarray}\label{eq:machocal}
V&=&V_{\mathrm{MACHO}}+24.22~\mathrm{mag}-0.18(V_{\mathrm{MACHO}}-R_{\mathrm{MACHO}})
\nonumber \\
R&=&R_{\mathrm{MACHO}}+23.98~\mathrm{mag}+0.18(V_{\mathrm{MACHO}}-R_{\mathrm{MACHO}}).
\end{eqnarray}
and the following one for the SMC:
\begin{eqnarray}\label{eq:smcmachocal}
V&=&V_{\mathrm{MACHO}}+24.97~\mathrm{mag}-0.18(V_{\mathrm{MACHO}}-R_{\mathrm{MACHO}})
\nonumber \\
R&=&R_{\mathrm{MACHO}}+24.73~\mathrm{mag}+0.18(V_{\mathrm{MACHO}}-R_{\mathrm{MACHO}}).
\end{eqnarray}
The zero point in Eq. \ref{eq:smcmachocal} is different from the one in Eq. \ref{eq:machocal} because of the different exposure times for LMC and SMC \citep{alcock99}.
\par
There are several hundreds of observations in both band passes for most EBs; the central fields of the LMC were observed more frequently than the periphery and there are on average fewer observations in the red band because one half of one of the red CCDs was out of commission during part of the project.
\par
For each source in the database found to be variable \citep{cook95} a period was found using the Supersmoother algorithm \citep[][first published by \citet{fri84}]{reimann94}.
The algorithm folds the unfolded light curve around different trial periods and selects the one that gives the smoothest light curve.
Periods were found separately for the red and blue unfolded light curve and they usually agree to high accuracy ($0.03\%$ on average); unfolded light curves were then folded around these periods \citep{faccioli07}.
\subsection{OGLE-II data}
The OGLE-II data we considered comprise the LMC sample of $2580$ EBs described in \citet{wyr03} 
of which $1182$ were also present in our sample, and the SMC sample of $1351$ EBs described in 
\citet{wyr04} of which $677$ were also present in our sample.
The available data include the phase of secondary minimum $\phi_\mathrm{sec}$
(the phase of primary minimum is set to $0$), the period, and $B$, $V$, $I$, and $I_\mathrm{DIA}$ magnitudes at maximum light.
Here $B$, $V$, and $I$ refer to magnitudes obtained using standard PSF fitting photometry
\citep{szymanski05} and $I_\mathrm{DIA}$ are $I$ band magnitudes computed via Difference Image Analysis photometry \citep[DIA:][]{zebrun01b,szymanski05}.
We used both $I$ and $I_\mathrm{DIA}$ magnitudes in our analysis.
The phase of the secondary $\phi_\mathrm{sec}$ is given up to two decimal places.
\section{Light curve fitting}
\label{sec:fit}
\begin{figure}
\footnotesize
\begin{center}
\plottwo{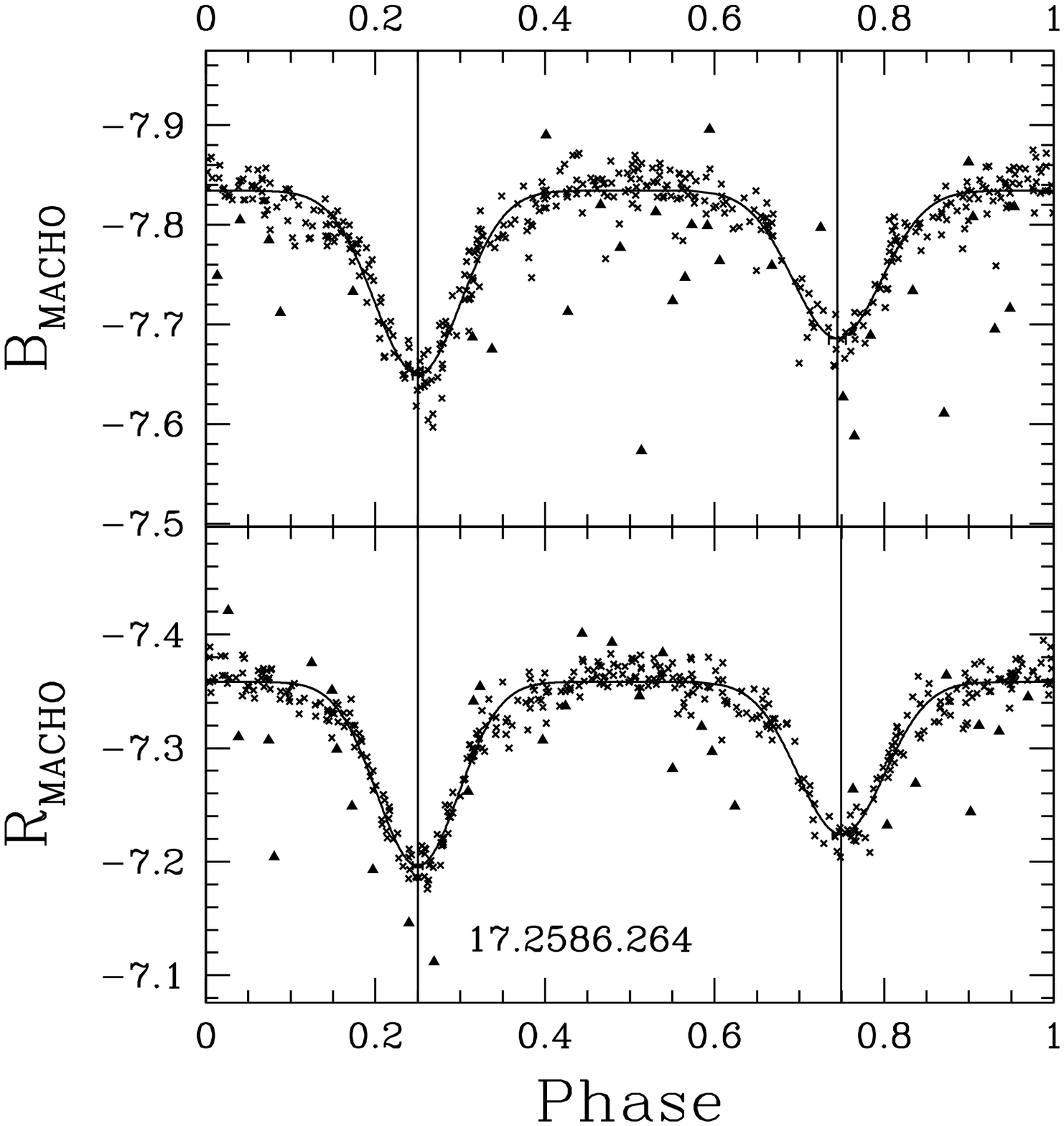}{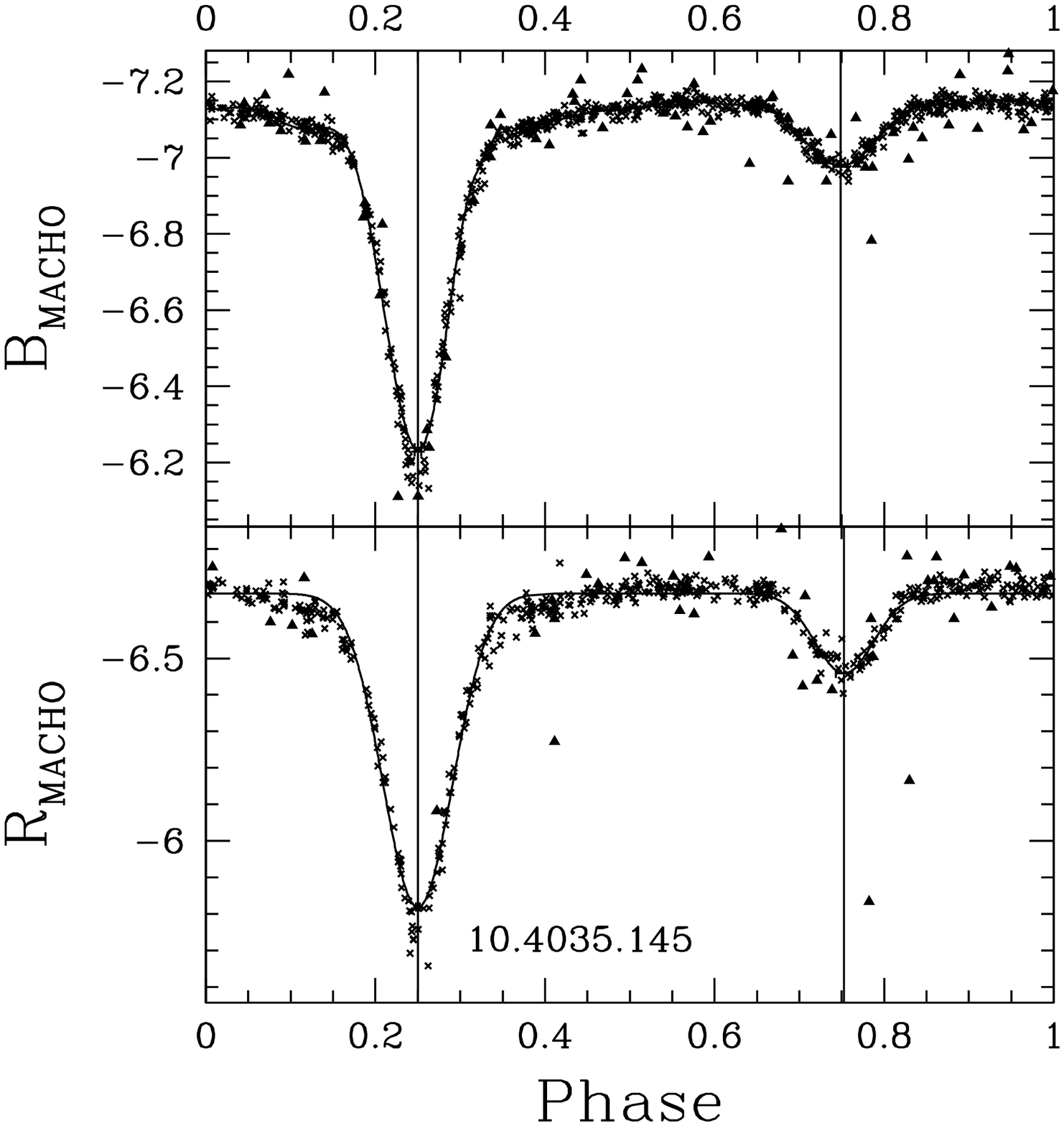}
\plottwo{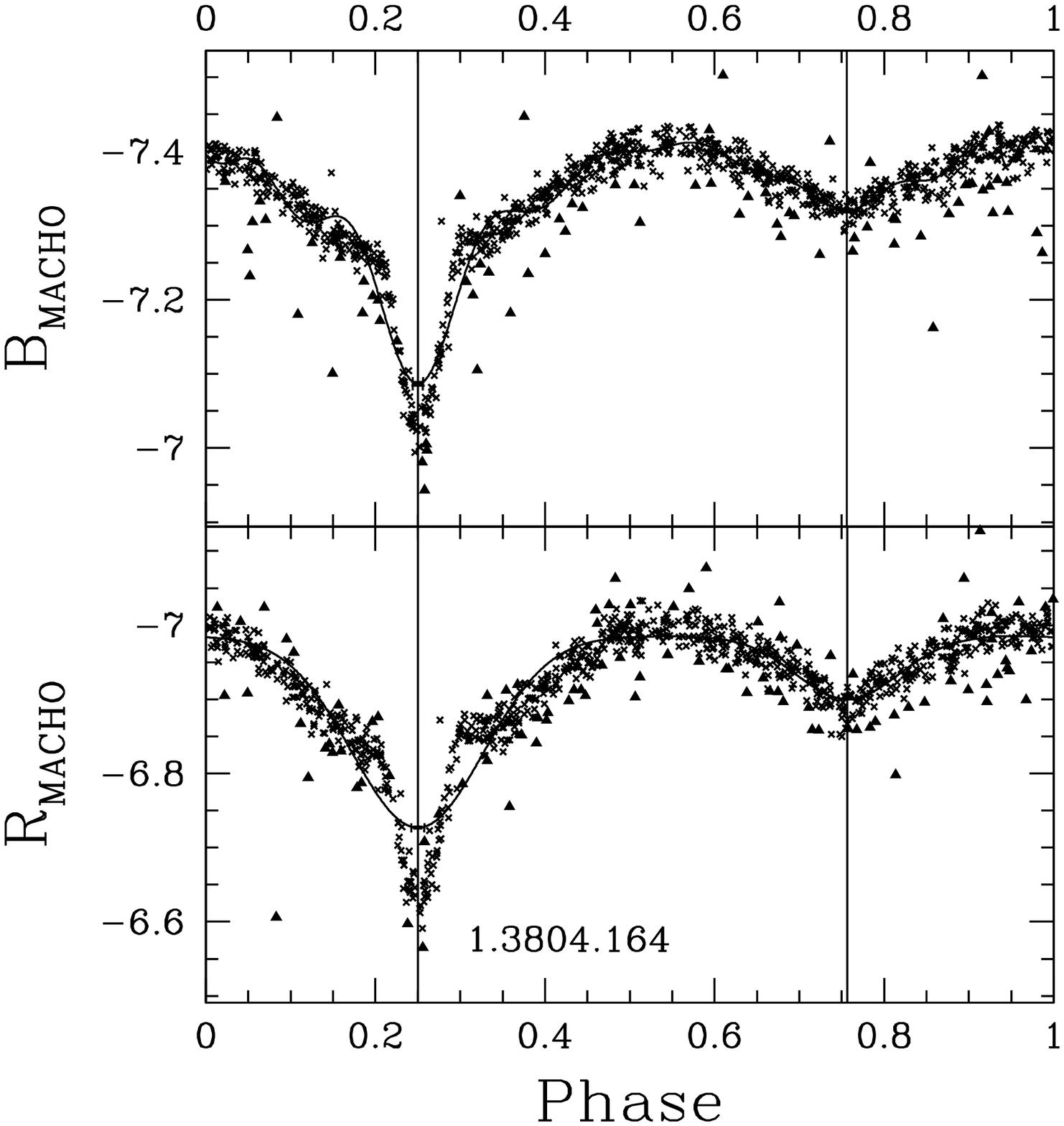}{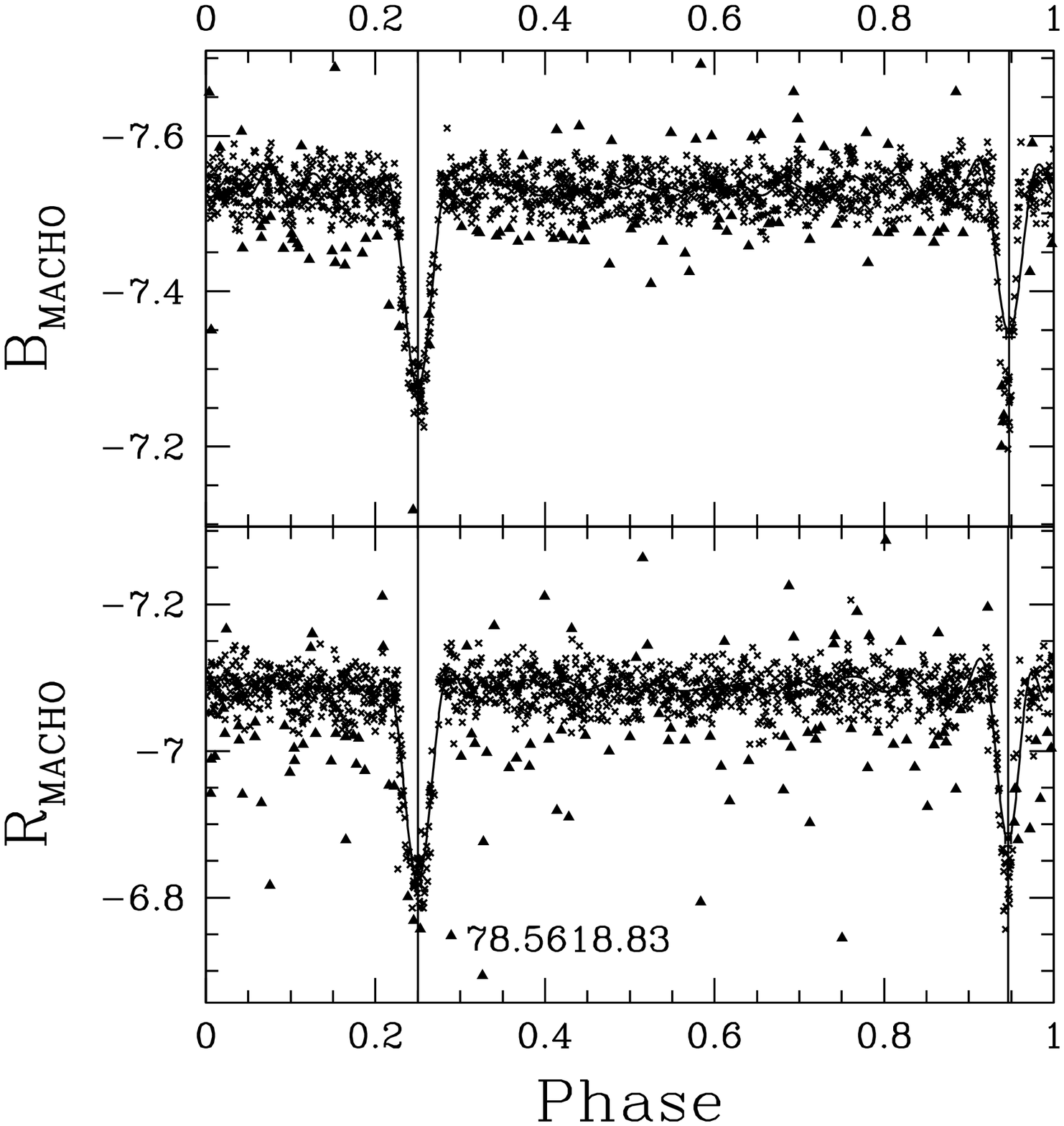}
\caption{Examples of light curve fitting with trigonometric functions used for minimum determinations for some LMC EBs in the MACHO sample.
Triangular points indicate outlying points excluded from the fits.
The light curves are shown in order of increasing period.}
\label{fig:sinlc}
\end{center}
\end{figure}

To find the phases of minima for the MACHO samples the light curves have been fitted to a sum of sines and cosines:
\begin{equation}\label{eq:sinfit}
m(\phi)=\Sigma_j(a_j\cos\omega_j\phi+b_j\sin\omega_j\phi)
\end{equation}
where $m(\phi)$ are the instrumental MACHO magnitudes, $V_{\mathrm{MACHO}}$ and $R_{\mathrm{MACHO}}$ in which the fits have been carried out and $\phi$ is the orbital phase for the light curves.
We preferred instrumental MACHO magnitudes because in many cases observations in \emph{one} of the MACHO bands are invalid, thus forcing us to exclude these points in the fit for \emph{both} bands if standard magnitudes had been used.
The ``frequencies'' $\omega_j$ have been found with the Lomb Periodogram technique \citep{lomb76,scargle82,press92}; this method is not a Fourier decomposition because the frequencies are not in a harmonic series; furthermore, the data are not uniformly spaced in phase.
Before fitting, outlying points in the light curves have been eliminated using moving windows with roughly $\sim N_{\mathrm{TOT}}/50$ (where $N_{\mathrm{TOT}}$ is the number of points in the light curve) points, calculating their mean and standard deviation and excluding from the fit points that are $>2$ standard deviations away from the mean in each window.
Figure \ref{fig:sinlc} shows some examples of sinusoidal fits and data on these objects are shown in Table \ref{tab:lc}.
The phasing was performed using the red and blue periods for the red and blue bands respectively.
The number of frequencies calculated by the Lomb Periodogram varies from object to object, but in most cases is in the range $10-100$.
\normalsize
\par
\begin{figure}
\footnotesize
\begin{center}
\plottwo{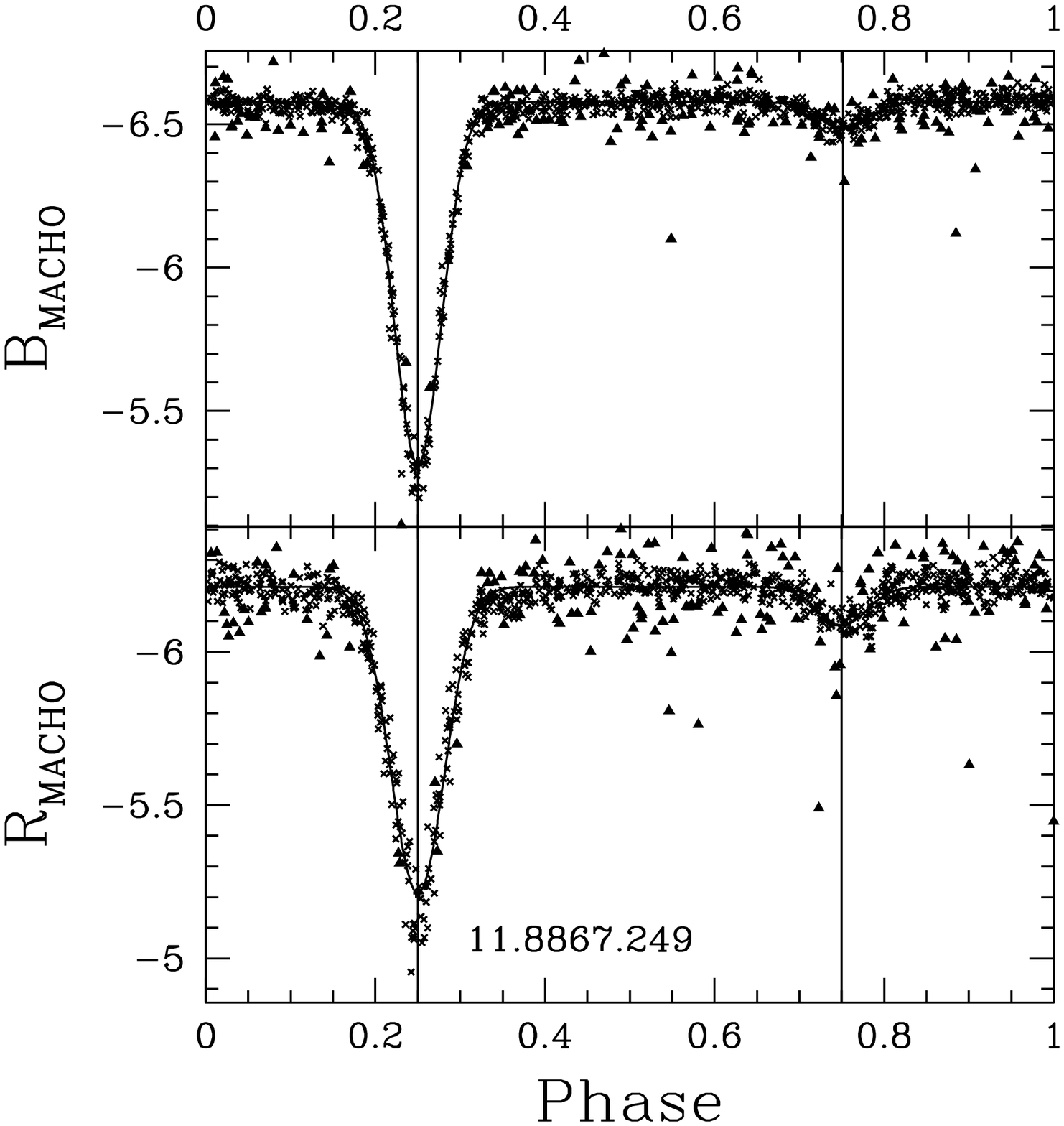}{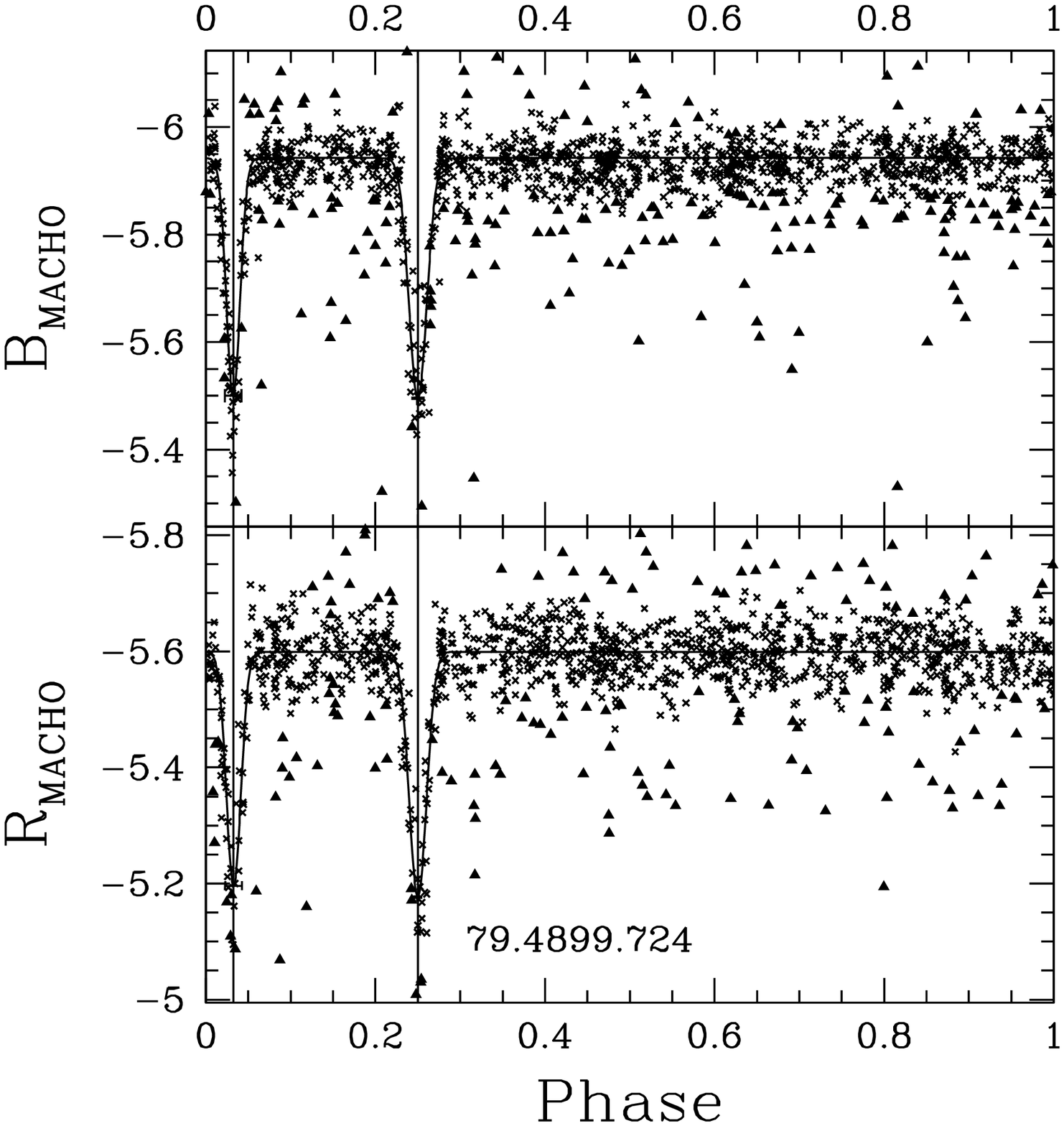}
\plottwo{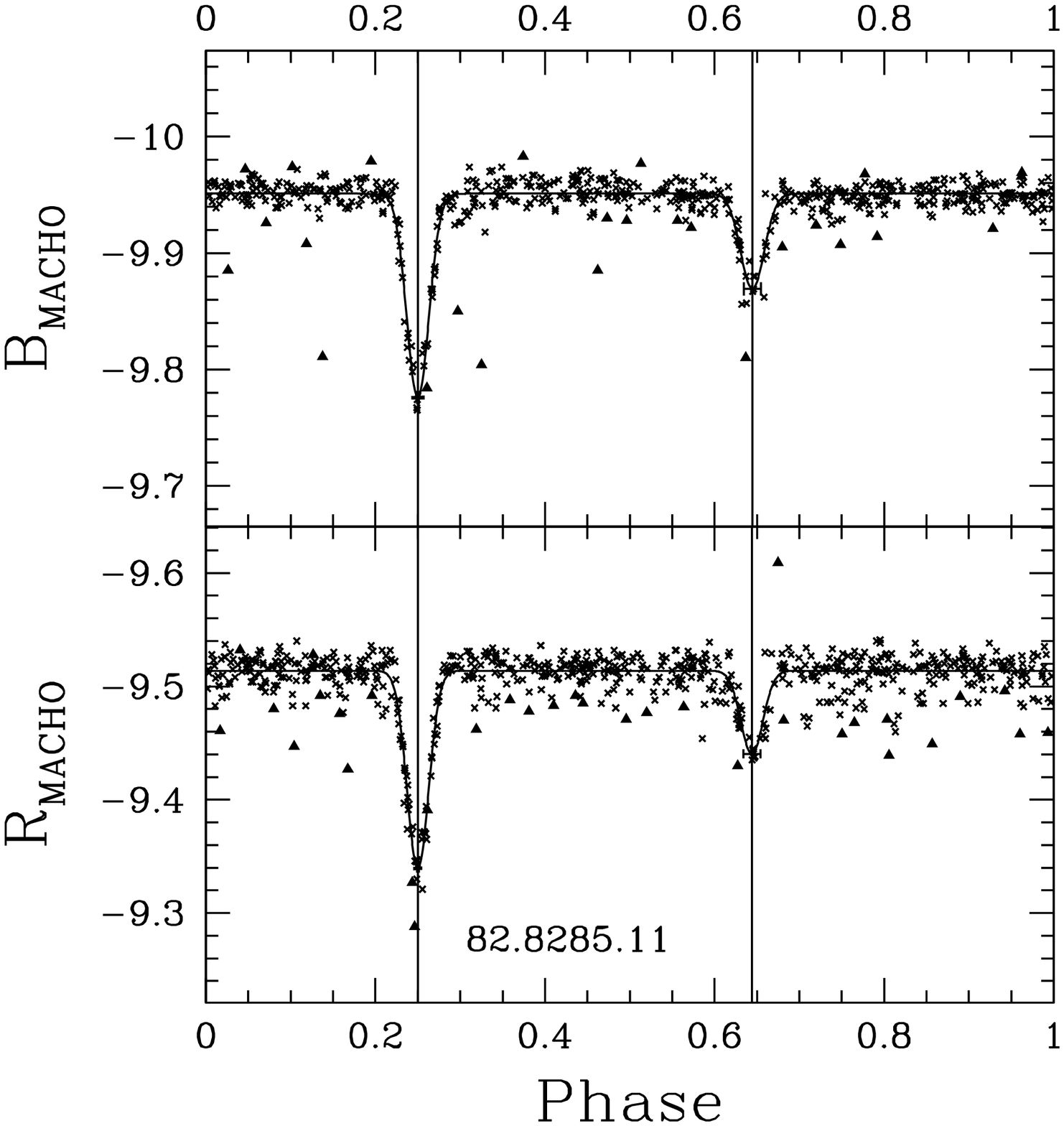}{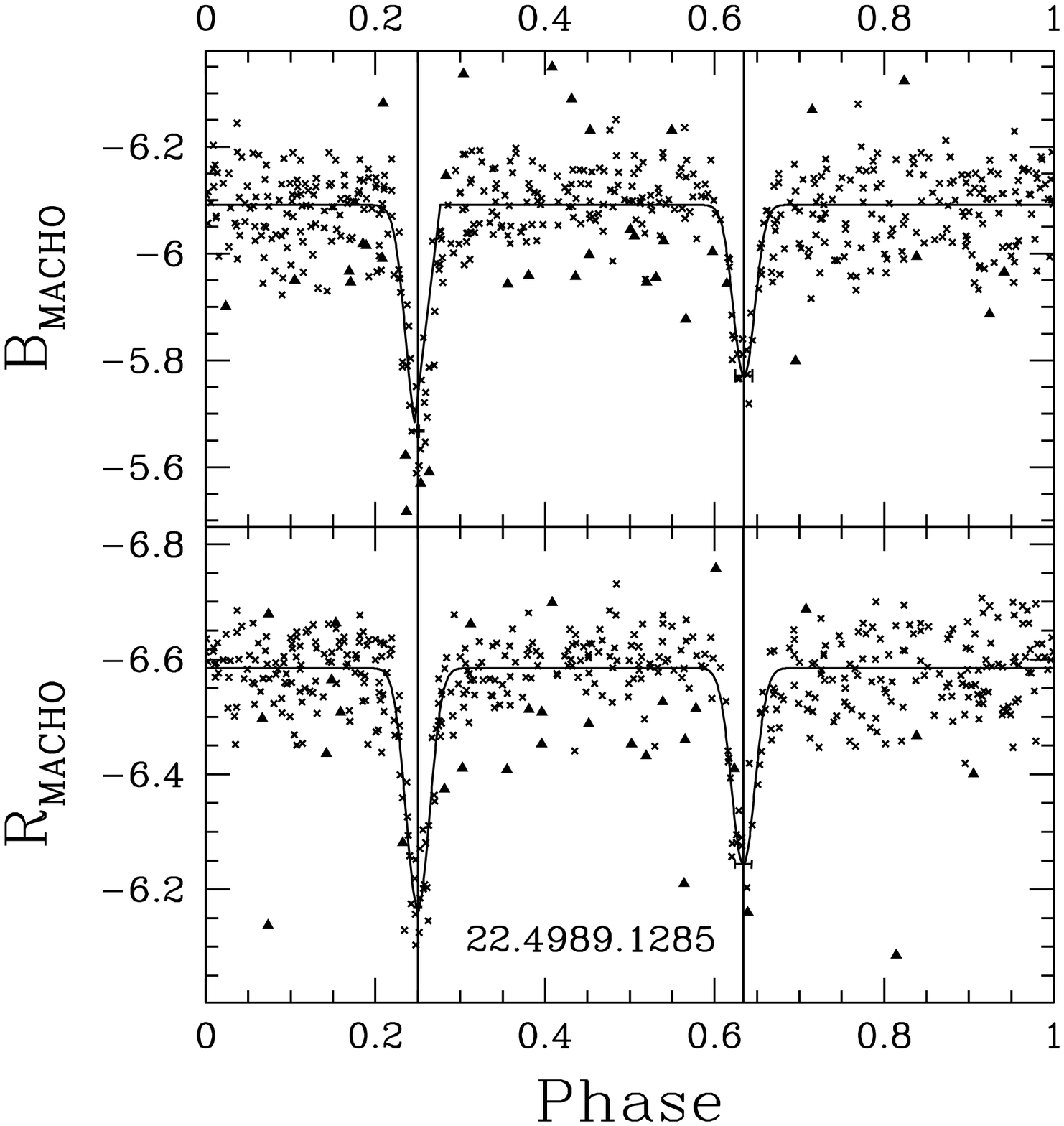}
\caption{Example of light curve fitting with Gaussians used for minimum 
determinations for some EBs in the LMC.
Triangular points indicate outlying points eliminated from the fit.
The light curves are shown in order of increasing period.}
\label{fig:gausslc}
\end{center}
\end{figure}

One problem of sinusoidal fit is that it tends to underestimate the depth of the minima for deep, widely separated eclipses; in general this fit works better for tidally distorted systems and less well for widely detached systems.
Therefore, after performing the sinusoidal fit, the light curves were also fitted to a sum of two Gaussians, which often better represent detached systems.
We used the positions of the minima determined via sinusoidal fit as a starting point for the Gaussians which was performed via the Levenberg-Marquardt algorithm \citep{press92}.
The best of the two fits from either sinusoidal or Gaussian fit for each light curve in each band was then selected.
Figure \ref{fig:gausslc} shows some examples of Gaussian fits; data on these objects are shown in Table \ref{tab:lc}.
\normalsize
\tabletypesize{\footnotesize}
\begin{deluxetable}{cccccccc}
\tablecolumns{4}
\tablewidth{0pc}
\tablecaption{Basic data for LMC EBs with light curves shown in Figures \ref{fig:sinlc} and \ref{fig:gausslc}.
The light curves are shown in order of increasing period.
\label{tab:lc}}
\tablehead{
\colhead{MACHO ID} & 
\colhead{RA(J2000)} & 
\colhead{DEC(J2000)} & 
\colhead{$P(\mathrm{d})$} &
\colhead{$V$\tablenotemark{a}} &
\colhead{$\vr$\tablenotemark{a}} &
\colhead{$\phi_1-\phi_2$} &
\colhead{Type of fit}
}
\startdata
17.2586.264  & 04:56:08.462 & -69:56:34.48 & 1.56 & 16.50 & -0.06 & 0.50 & 
Sinusoidal \\
10.4035.145  & 05:05:02.233 & -70:06:13.27 & 2.53 & 17.24 & -0.05 & 0.55 & 
Sinusoidal \\
11.8867.249  & 05:35:14.688 & -70:40:19.12 & 3.14 & 17.72 &  0.01 & 0.50 & 
Gaussian \\
1.3804.164   & 05:03:36.536 & -69:23:32.27 & 4.19 & 17.20 & -0.02 & 0.50 & 
Sinusoidal \\
79.4899.724  & 05:10:21.705 & -69:01:01.48 & 5.00 & 18.05 &  0.03 & 0.62 & 
Gaussian \\
82.8285.11   & 05:31:04.061 & -69:09:20.81 & 11.61 & 16.85 & -0.04 & 0.50 &
Gaussian \\
78.5618.83   & 05:15:19.305 & -69:26:39.97 & 15.96 & 14.35 & -0.04 & 0.61 &
Sinusoidal \\
22.4989.1285 & 05:11:31.896 & -71:01:45.95 & 107.24 & 18.28 & 0.55 & 0.74 &
Gaussian \\
\enddata
\tablenotetext{a}{Values are quoted to the hundredths of magnitude, typical of MACHO observational uncertainties.
\tablenotetext{b}{Values are quoted to 1\%, typical of the uncertainties of our fitting procedure.}
}
\end{deluxetable}
\par
\begin{figure}
\footnotesize
\begin{center}
\plotone{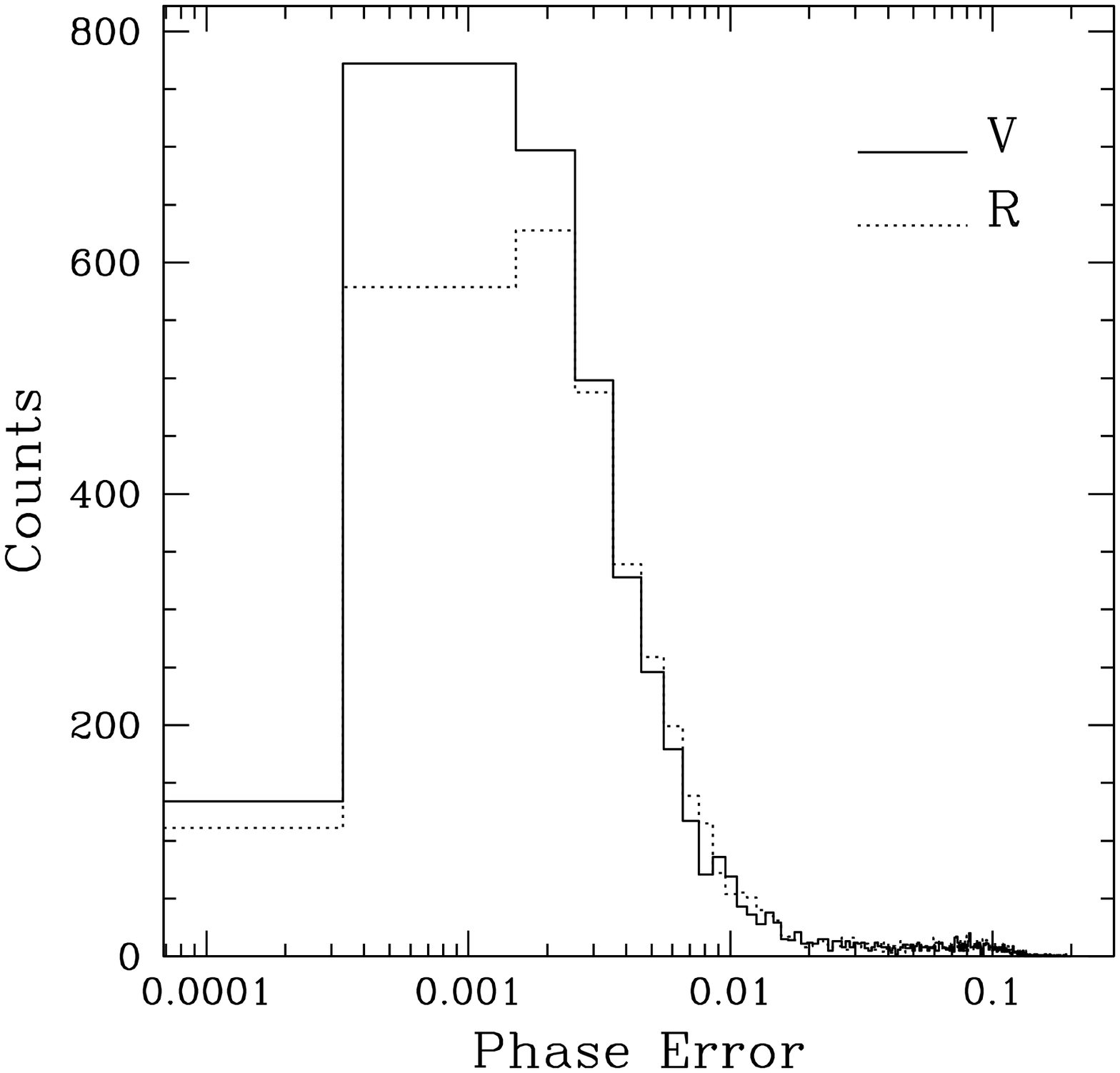}
\caption{Histograms of the errors in the phase differences of the minima from the Monte Carlo simulation.}
\label{fig:errhist}
\end{center}
\end{figure}
\begin{figure}
\footnotesize
\begin{center}
\plotone{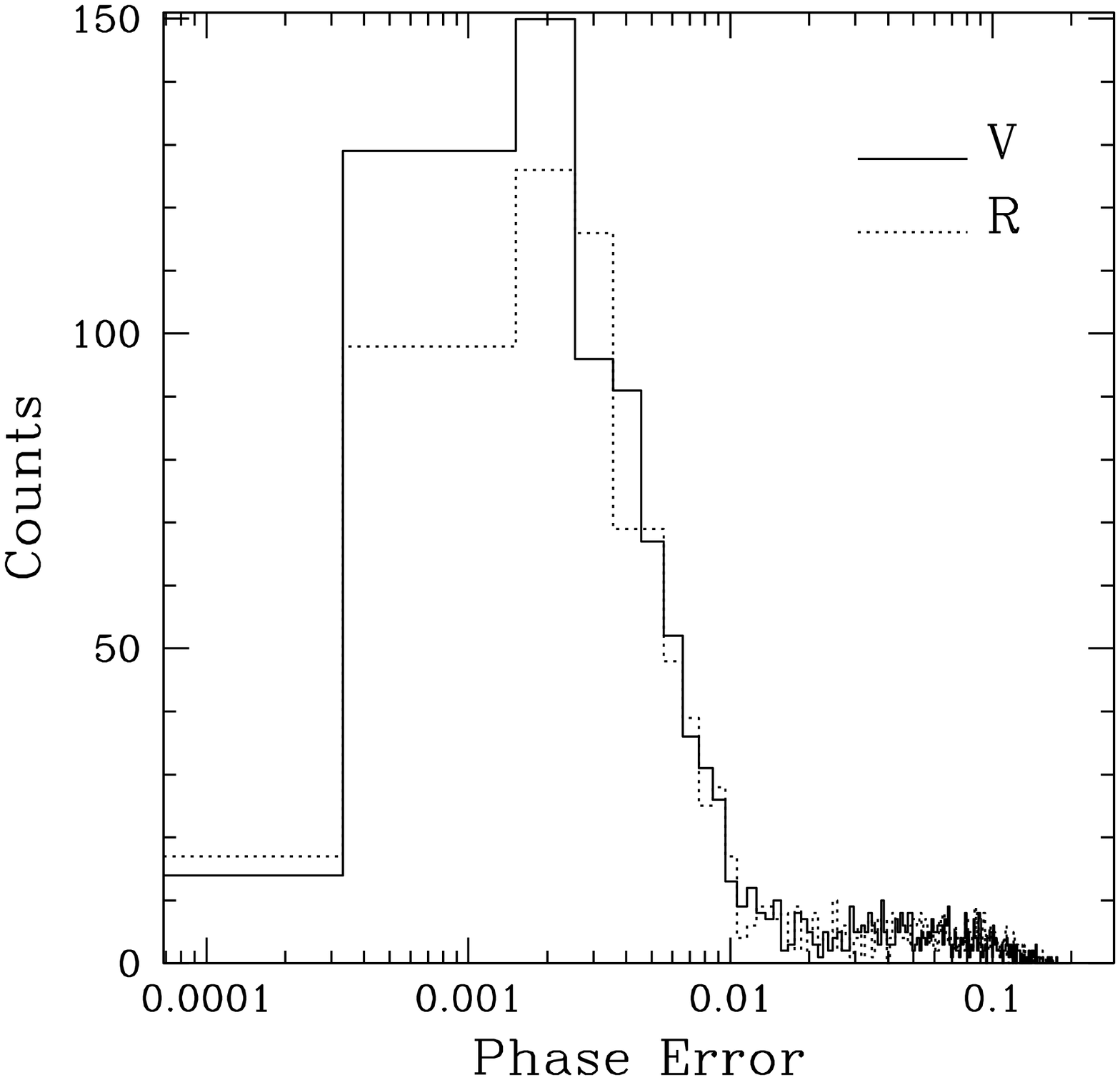}
\caption{Histograms of the errors in the phase differences of the minima from the Monte Carlo simulation with a random shift added to the phases.}
\label{fig:errhistrandomphase}
\end{center}
\end{figure}

It is necessary to estimate the errors in our determinations of the quantity $\phi_1-\phi_2$.
We did this as follows for each of our systems.
First, we created a synthetic light curve using the EBOP code \citep{etzel81,popper81}, which implements the model by \citet{nelson72} with some modifications and we followed the prescriptions described in \citet{lacy97}.
This synthetic light curve was used only to test our procedure for estimating $\phi_1-\phi_2$.
and was sampled at the same phases $\phi_i$ as the original light curve; noise was added to mimic a MACHO light curve.
The synthetic noise was determined from the EBOP fit to the MACHO light curve as follows.
First the residuals $O_i-C_i$ were determined, where $O_i$ is the observed magnitude at phase $\phi_i$ and $C_i$ the value of the fitted magnitude at that phase.
These residuals were divided by the MACHO estimated photometric errors $\sigma_i$ to create scaled residuals $s_i=\frac{O_i-C_i}{\sigma_i}$.
The cumulative distribution of the $s_j$ was sampled randomly and used to create the synthetic noise at phase $\phi_i$, $N_i=s_j\sigma_i$.
The new, synthetic, noisy light curve was then $O_{i}^{s}=C_i+N_i$.
We determined the quantity $\phi_1-\phi_2$ for this light curve, and the entire procedure was repeated $\sim$~30 times in each band.
We excluded the cases in which $\phi_1-\phi_2>0.9$ (as we did in the real fits) 
which we took as an indication that the fitting procedure had failed.
The RMS of the deviations from $\phi_1-\phi_2$ determined from the MACHO data is a reasonable estimate of the error in this procedure.
Figure \ref{fig:errhist} shows the histograms of the errors thus defined for both bands; most errors are smaller than $0.01$ making our method for detecting eccentricities very robust.
\normalsize
We then studied the effects of period uncertainties on our method.
A small period uncertainty should manifest itself as a constant shift in the 
phases of a light curve: therefore we selected a random subsample of $\sim 50$ 
EBs, created simulated light curves in the same way but in addition we added a 
constant random phase shift to them and repeated the procedure.
The results are shown in Figure \ref{fig:errhistrandomphase} which shows that our
method is also robust with respect to uncertainties in the period determination.
\par
In the few cases where our procedure gives a large error estimate the reason
is mostly the presence of one or two shallow minima and/or noise in the light curve.
In the presence of a very shallow eclipse a small perturbation of the data can lead to a 
large difference in the determination of one or both phases of the minima and
therefore a large variance used to estimate the error in phase difference; this is true
also for noisy data.
\normalsize
\section{Results}
\label{sec:results}
\subsection{LMC results from MACHO}
\begin{figure}
\footnotesize
\begin{center}
\plotone{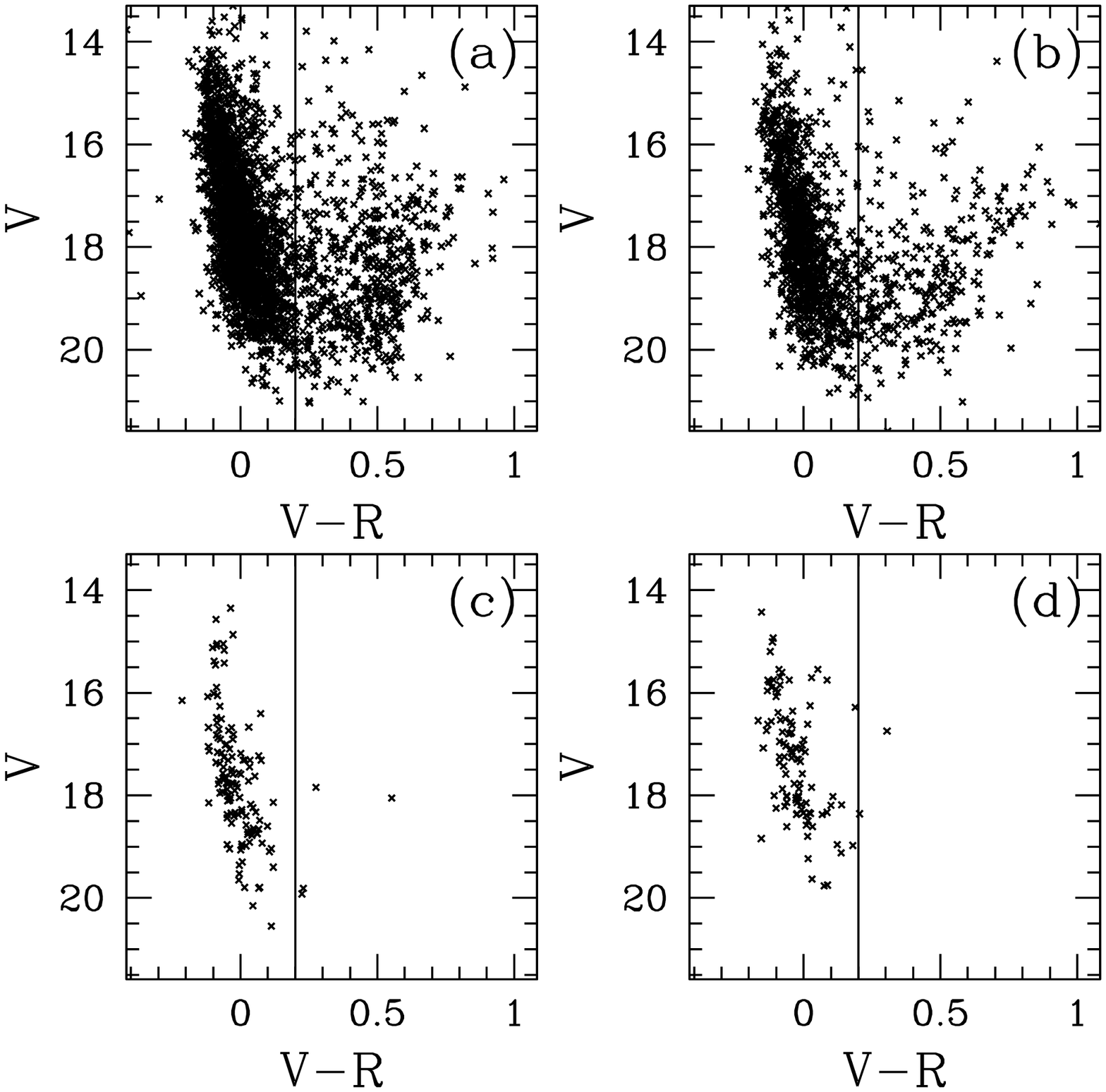}
\caption{Color-Magnitude Diagram for $4510$ EBs in the LMC MACHO sample.
Panel (a): $|\phi_2-\phi_1|<0.51$,
Panel (b): $0.51<|\phi_2-\phi_1|<0.6$,
Panel (c): $0.6<|\phi_2-\phi_1|<0.65$,
Panel (d): $|\phi_2-\phi_1|>0.65$.
}
\label{fig:cmd}
\end{center}
\end{figure}

\begin{figure}
\footnotesize
\begin{center}
\plotone{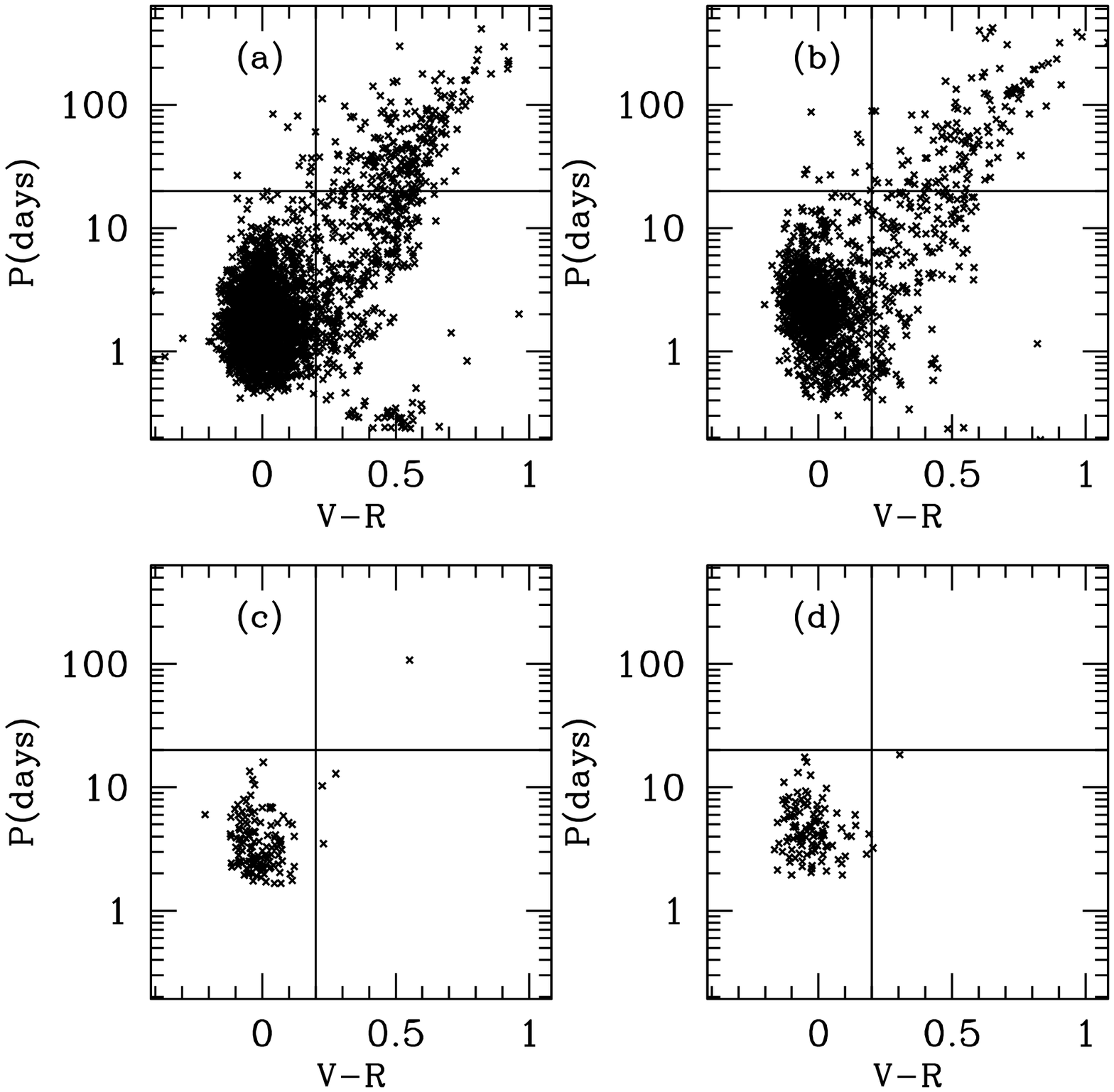}
\caption{Color-Period Diagram for $4510$ EBs in the LMC MACHO sample.
Panel (a): $|\phi_2-\phi_1|<0.51$,
Panel (b): $0.51<|\phi_2-\phi_1|<0.6$,
Panel (c): $0.6<|\phi_2-\phi_1|<0.65$,
Panel (d): $|\phi_2-\phi_1|>0.65$.}
\label{fig:percol}
\end{center}
\end{figure}

For the LMC our fitting procedure gave acceptable results in at least one band for $4510$ EBs out
of the $4634$ EBs that make up the LMC sample described in \citet{faccioli07}. 
Figures \ref{fig:cmd} and \ref{fig:percol} report the Color Magnitude Diagram and the Color Period Diagram respectively for the LMC EBs in our sample in different panels according to their phase difference.
The explanation of the panels is given in Table \ref{tab:panel}, which also reports the minimum eccentricity corresponding to a given phase difference, obtained by assuming $\cos\omega=1$ in Eq. \ref{eq:ecc}.
For the purpose of this discussion we describe EBs with $\vr<0.2~\mathrm{mag}$ as Main Sequence systems and EBs with $\vr>0.2~\mathrm{mag}$ as evolved systems; we also describe EBs with $P>20\mathrm{d}$ as ``long period'' EBs; the breakup of MACHO LMC and SMC EBs and OGLE-II LMC EBs used in our analysis is given by Table \ref{tab:smcvslmc}.
\tabletypesize{\footnotesize}
\begin{deluxetable}{ccc}
\tablecolumns{3}
\tablewidth{0pc}
\tablecaption{Explanation of panel labels for Figures \ref{fig:cmd}, \ref{fig:percol},
\ref{fig:cmdogle}, \ref{fig:percologle}, \ref{fig:cmd.smc}, \ref{fig:percol.smc}
\ref{fig:cmdoglesmc}, and \ref{fig:percologlesmc}.
\label{tab:panel}}
\tablehead{
\colhead{Panel label} & 
\colhead{Phase difference} &
\colhead{Minimum eccentricity}
}
\startdata
(a) & $|\phi_2-\phi_1|<0.51$ & $e>0$ \\
(b) & $0.51<|\phi_2-\phi_1|<0.6$ & $e>0.016$ \\
(c) & $0.6<|\phi_2-\phi_1|<0.65$ & $e>0.16$ \\
(d) & $|\phi_2-\phi_1|>0.65$ & $e>0.24$ \\
\enddata
\end{deluxetable}
\tabletypesize{\footnotesize}
\begin{deluxetable}{cccccccc}
\tablecolumns{8}
\tablewidth{0pc}
\tablecaption{Summary of the sample features.
\label{tab:smcvslmc}}
\tablehead{
\colhead{Galaxy} &
\colhead{Sample} &
\colhead{Total} &
\colhead{MS\tablenotemark{a}} &
\colhead{Evolved\tablenotemark{b}} &
\colhead{Long Period\tablenotemark{c}} &
\colhead{Long Period\tablenotemark{c} MS\tablenotemark{a}} &
\colhead{Long Period\tablenotemark{c} Evolved\tablenotemark{b}}
}
\startdata
LMC &
MACHO &
$4510$ &
$3667 (81\%)$ &
$843 (19\%)$ &
$349$ &
$21 (6\%)$ &
$328 (94\%)$ \\
LMC &
OGLE-II &
$2474$ &
$1744 (70\%)$ &
$730 (30\%)$ &
$216$ &
$31 (14.3\%)$ &
$185 (85.7\%)$ \\
SMC &
MACHO & 
$1380$ &
$1293 (94\%)$ &
$87 (6\%)$ &
$66$ &
$19 (29\%)$ &
$47 (71\%)$ \\
SMC &
OGLE-II &
$1317$ &
$1091 (83\%)$ &
$226 (17\%)$ &
$165$ &
$27 (16.4\%)$ &
$138 (83.6\%)$ \\
\enddata
\tablenotetext{a}{Main Sequence: defined as $\vr<0.2~\mathrm{mag}$ for the MACHO samples and as 
$V-I_{\mathrm{DIA}}<0.4~\mathrm{mag}$ for the OGLE-II samples.}
\tablenotetext{b}{Defined as $\vr>0.2~\mathrm{mag}$ for the MACHO samples and as 
$V-I_{\mathrm{DIA}}>0.4~\mathrm{mag}$ for the OGLE-II samples.}
\tablenotetext{c}{Defined as $P>20\mathrm{d}$ both for the MACHO and the OGLE-II 
samples.}
\end{deluxetable}
It is evident from the Color Magnitude Diagram (Figure \ref{fig:cmd}) that evolved systems have $|\phi_2-\phi_1|<0.6$ in the vast majority of cases; indeed there is a cutoff in phase difference at $\vr\sim 0.2~\mathrm{mag}$.
\normalsize
\normalsize
The Color Period Diagram shown in Figure \ref{fig:percol} shows the cutoff in $\vr$ too, and also shows that elliptical orbits are concentrated in the 
$1.5\mathrm{d}\lesssim P\lesssim 20\mathrm{d}$ range.
This is not surprising in view of Kepler's third law $P\propto a^{3/2}$,where $a$ is the semi-major axis: in a system with long period and hence widely separated stars, tidal forces are less effective at orbital circularization and orbits remain elliptical longer.
Conversely in systems with shorter period, tidal forces can achieve orbital circularization already on the Main Sequence.
The diagram shows a strong Color-Period relation for evolved, long period systems.
These systems are mostly ellipsoidal, with continuously varying light curves \citep{faccioli07}; a minority however is composed of detached systems or systems whose
eclipses, or at least the primary one, show a clear beginning and an end.
The small population ($\sim 60$ EBs) around $\vr\sim 0.5~\mathrm{mag}$ and $P\sim 0.3\mathrm{d}$ that is visible in Panel (a), and to a smaller extent in Panel (b), of Figure \ref{fig:percol} is probably due to foreground contamination.
We argue in \citep{faccioli07} that the period and color of these systems is compatible with them being composed of Main Sequence solar mass stars in the Milky Way; the fact that we do not see systems with highly eccentric orbit in this foreground population can be attributed to small number statistic.
\subsection{LMC results from OGLE-II}
\begin{figure}
\footnotesize
\begin{center}
\plotone{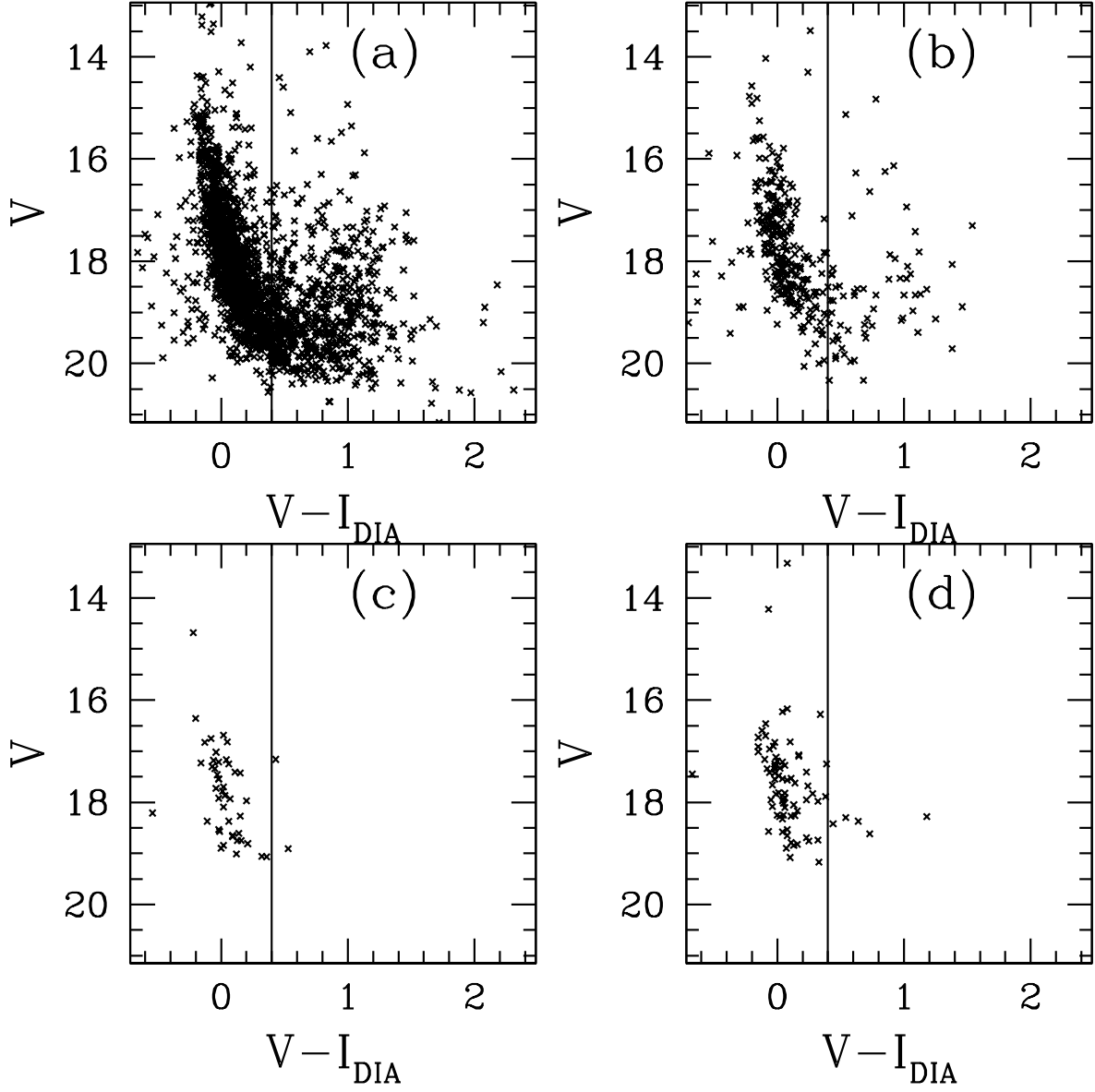}
\caption{Color-Magnitude Diagram for $2474$ EBs in the LMC OGLE-II sample.
Panel (a): $|\phi_2-\phi_1|<0.51$,
Panel (b): $0.51<|\phi_2-\phi_1|<0.6$,
Panel (c): $0.6<|\phi_2-\phi_1|<0.65$,
Panel (d): $|\phi_2-\phi_1|>0.65$.}
\label{fig:cmdogle}
\end{center}
\end{figure}

\begin{figure}
\footnotesize
\begin{center}
\plotone{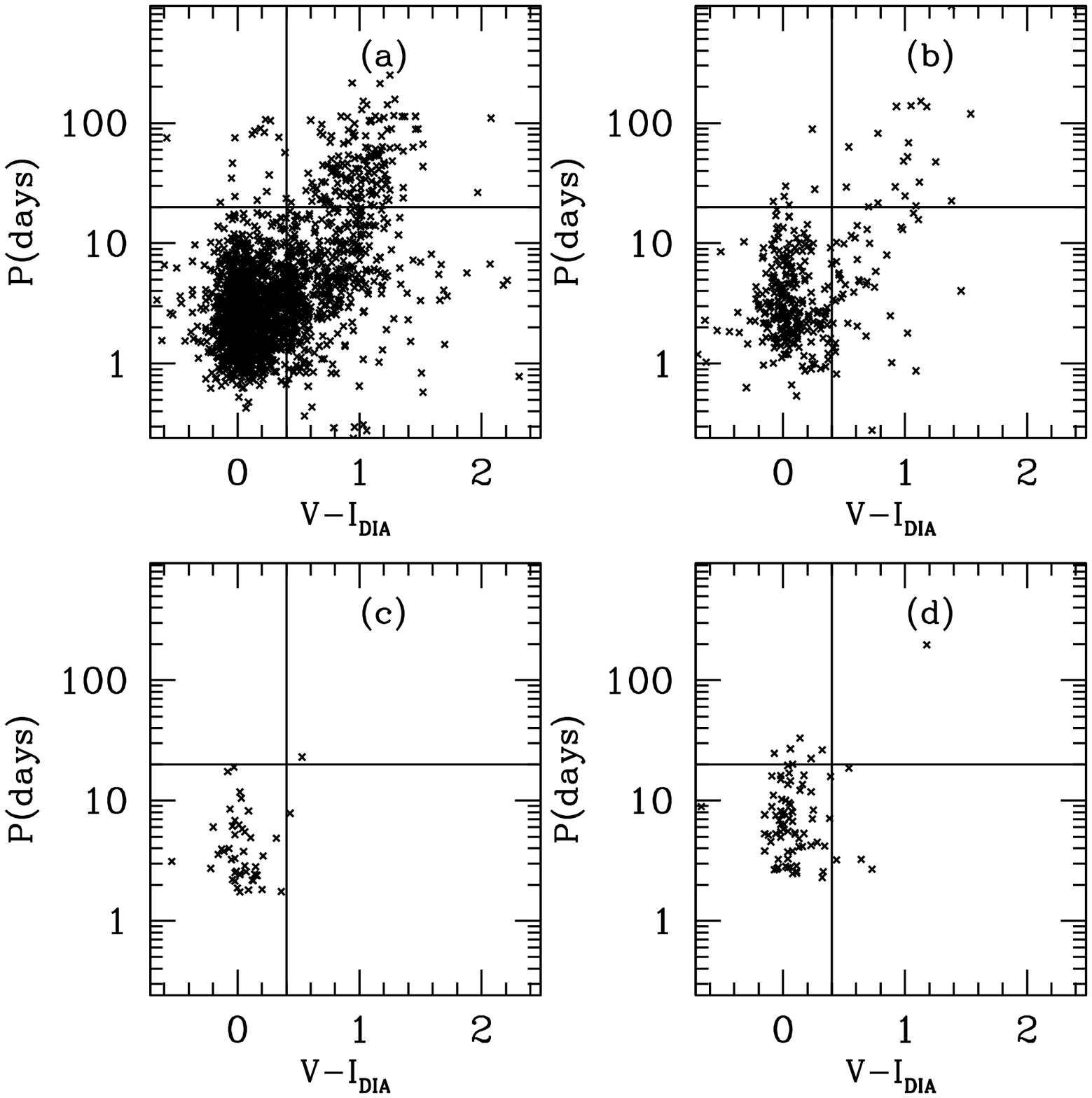}
\caption{Color-Period Diagram for $2474$ EBs in the LMC OGLE-II sample.
Panel (a): $|\phi_2-\phi_1|<0.51$,
Panel (b): $0.51<|\phi_2-\phi_1|<0.6$,
Panel (c): $0.6<|\phi_2-\phi_1|<0.65$,
Panel (d): $|\phi_2-\phi_1|>0.65$.}
\label{fig:percologle}
\end{center}
\end{figure}

From the $2580$ EBs in this sample we selected the $2525$ that had a valid $V$ magnitude and from these the $2474$ EBs for which a phase of secondary minimum was provided; since the phase of primary minimum is set to $0$, the phase of secondary minimum is equal to our $\phi_1-\phi_2$; when $\phi_\mathrm{sec}<0.5$ we just took $\phi_\mathrm{sec}\rightarrow 1-\phi_\mathrm{sec}$.
Figure \ref{fig:cmdogle} shows the Color Magnitude Diagram and Figure \ref{fig:percologle} shows the Color Period Diagram.
The $V-I_\mathrm{DIA}$ axis interval in both diagrams has been chosen so that it has roughly the same range as the $\vr$ axis in Figures. \ref{fig:cmd} and \ref{fig:percol} once the difference between $\vr$ and $V-I$ colors is taken into account; because of this cut $19$ OGLE-II EBs are not shown in
Figures \ref{fig:cmdogle} and \ref{fig:percologle}.
The basic features found in the MACHO sample are also evident in the OGLE-II sample, with evolved EBs having mostly circular orbits and EBs on the Main Sequence having a broad range of eccentricities.
\normalsize
\normalsize
\subsection{SMC results from MACHO and OGLE-II}
\begin{figure}
\footnotesize
\begin{center}
\plotone{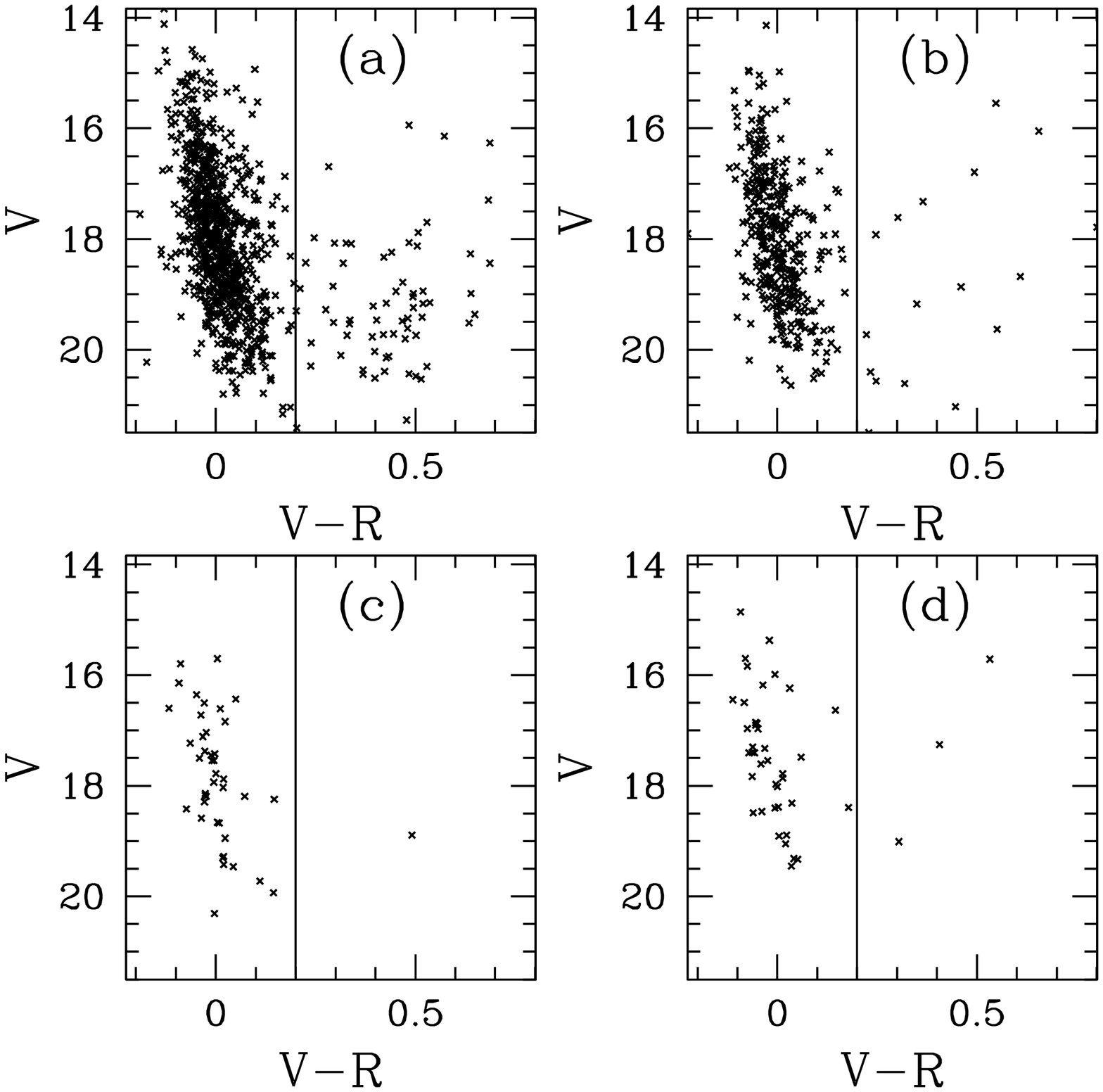}
\caption{Color-Magnitude Diagram for $1380$ EBs in SMC MACHO sample.
Panel (a): $|\phi_2-\phi_1|<0.51$,
Panel (b): $0.51<|\phi_2-\phi_1|<0.6$,
Panel (c): $0.6<|\phi_2-\phi_1|<0.65$,
Panel (d): $|\phi_2-\phi_1|>0.65$.}
\label{fig:cmd.smc}
\end{center}
\end{figure}

\begin{figure}
\footnotesize
\begin{center}
\plotone{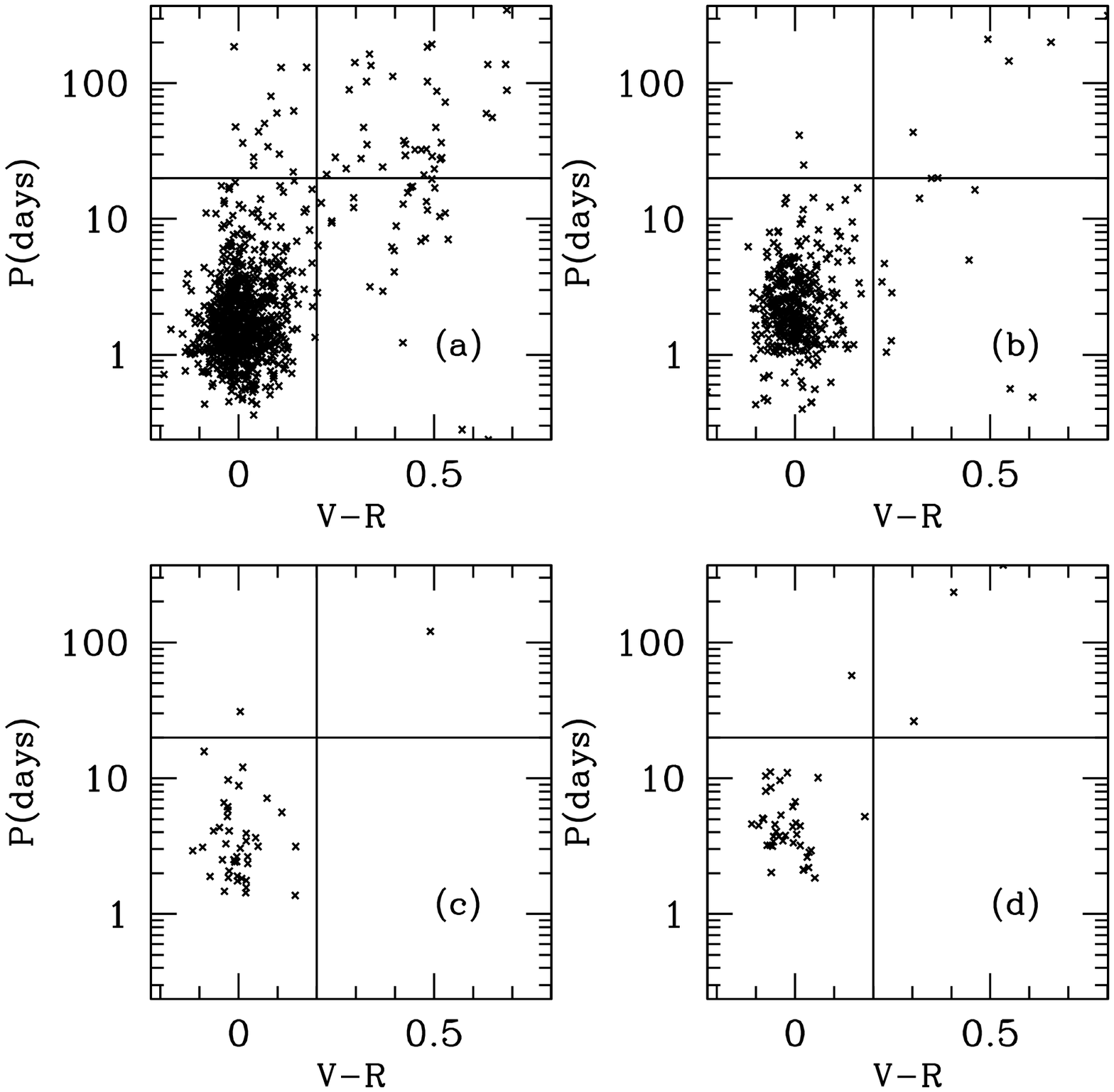}
\caption{Color-Period Diagram for $1380$ EBs in the SMC MACHO sample.
Panel (a): $|\phi_2-\phi_1|<0.51$,
Panel (b): $0.51<|\phi_2-\phi_1|<0.6$,
Panel (c): $0.6<|\phi_2-\phi_1|<0.65$,
Panel (d): $|\phi_2-\phi_1|>0.65$.}
\label{fig:percol.smc}
\end{center}
\end{figure}

For the SMC the fit, applied to the sample of $1509$ EBs introduced in \citep{faccioli07}, gave acceptable minima determinations in at least one band for $1380$ EBs.
We observe circularization past the Main Sequence in these systems as well as shown by Figures \ref{fig:cmd.smc} and \ref{fig:percol.smc}.
\normalsize
\normalsize

\begin{figure}
\footnotesize
\begin{center}
\plotone{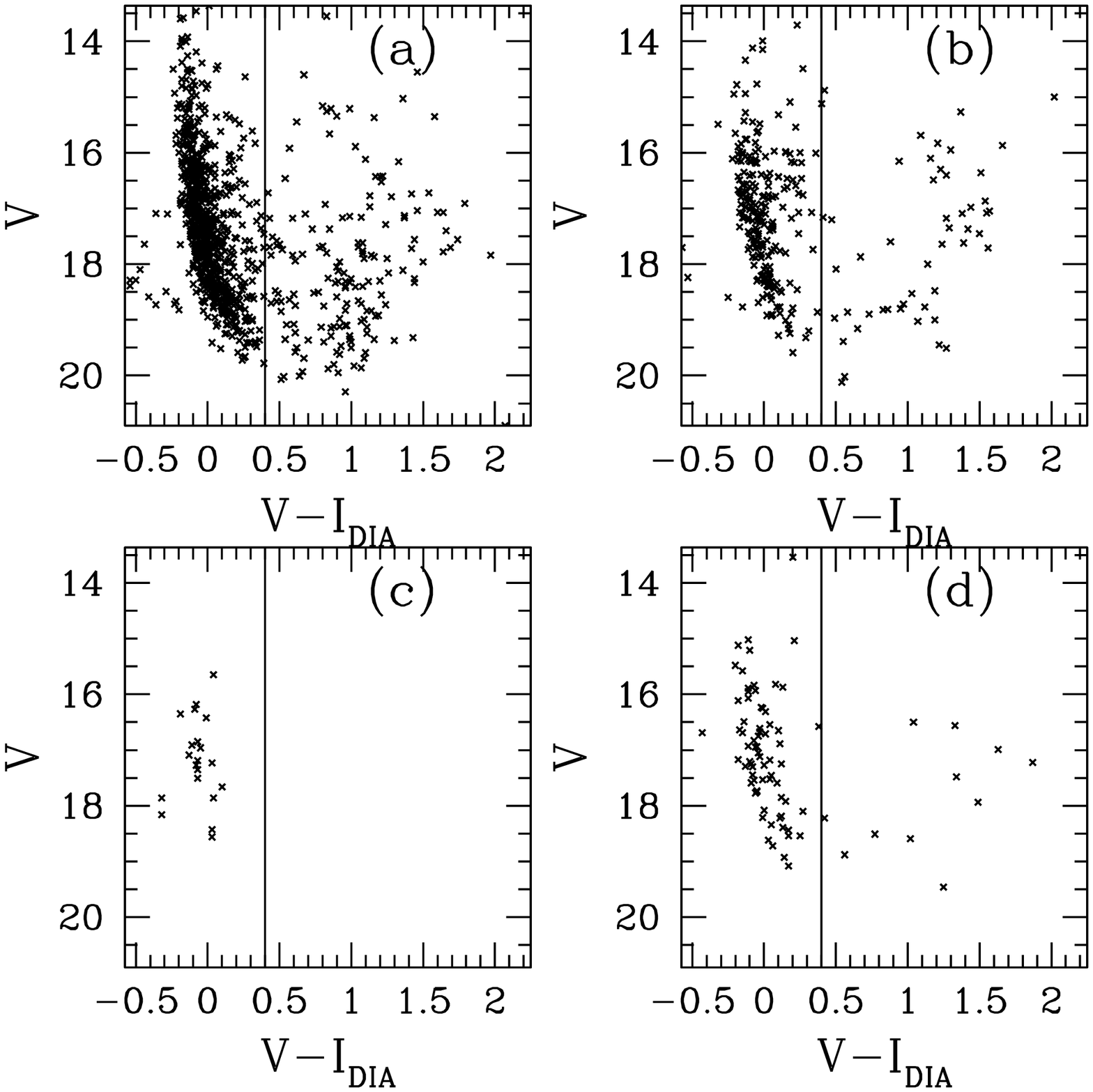}
\caption{Color-Magnitude Diagram for $1317$ EBs in the SMC OGLE-II sample.
Panel (a): $|\phi_2-\phi_1|<0.51$,
Panel (b): $0.51<|\phi_2-\phi_1|<0.6$,
Panel (c): $0.6<|\phi_2-\phi_1|<0.65$,
Panel (d): $|\phi_2-\phi_1|>0.65$.}
\label{fig:cmdoglesmc}
\end{center}
\end{figure}

\begin{figure}
\footnotesize
\begin{center}
\plotone{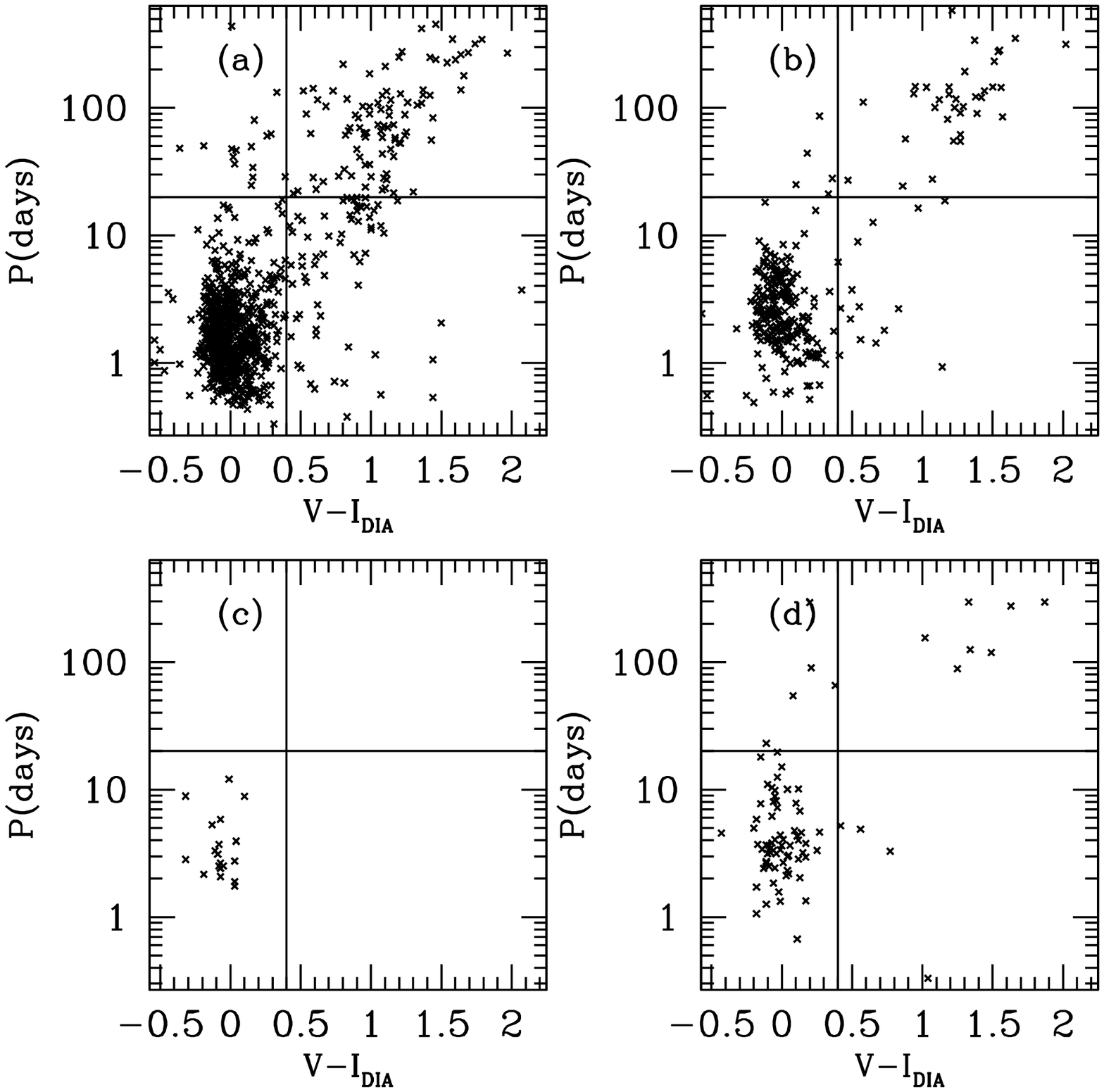}
\caption{Color-Period Diagram for $1317$ EBs in the SMC OGLE-II sample.
Panel (a): $|\phi_2-\phi_1|<0.51$,
Panel (b): $0.51<|\phi_2-\phi_1|<0.6$,
Panel (c): $0.6<|\phi_2-\phi_1|<0.65$,
Panel (d): $|\phi_2-\phi_1|>0.65$.}
\label{fig:percologlesmc}
\end{center}
\end{figure}

From the $1351$ EBs in the OGLE-II sample we selected the $1317$ that had a valid $V$ magnitude and for which a phase of secondary minimum was provided.
Figure \ref{fig:cmdoglesmc} shows the Color Magnitude Diagram and Figure \ref{fig:percologlesmc} shows the Color Period Diagram.
Again the $V-I_\mathrm{DIA}$ axis interval in both diagrams has been chosen so that it has roughly the same range as the $\vr$ axis in Figures. \ref{fig:cmd.smc} and \ref{fig:percol.smc} once the difference between $\vr$ and $V-I$ colors is taken into account; because of this cut $15$ OGLE-II EBs are not shown in Figures \ref{fig:cmdoglesmc} and \ref{fig:percologlesmc}.
The basic features found in the MACHO sample are also evident in the OGLE-II sample, with evolved EBs having mostly circular orbits and EBs on the Main Sequence having a broad range of eccentricities.
\normalsize
\normalsize
\subsection{Period-Phase Difference and Color-Phase Difference Diagrams}
\begin{figure}
\footnotesize
\begin{center}
\plottwo{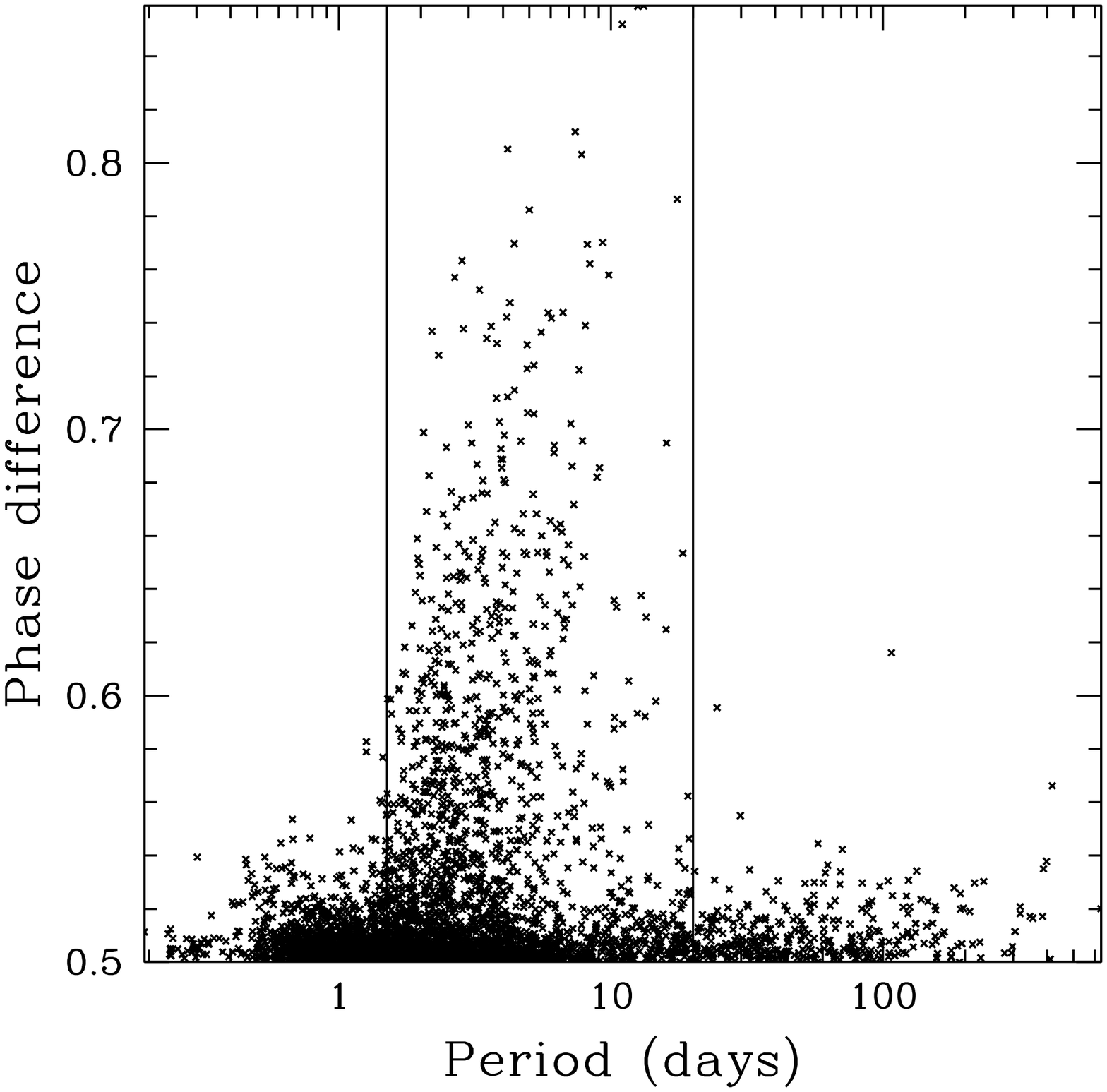}{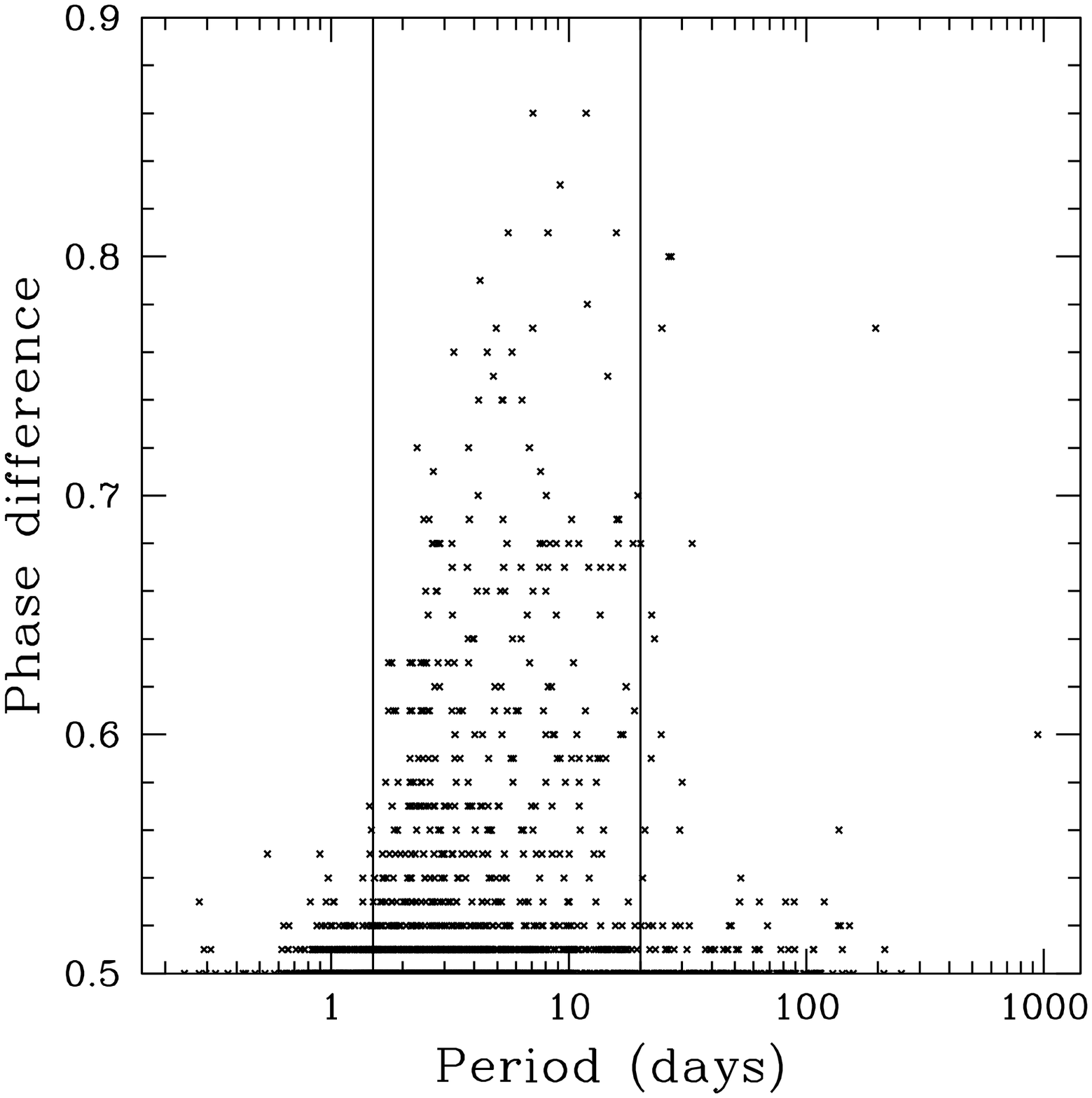}
\plottwo{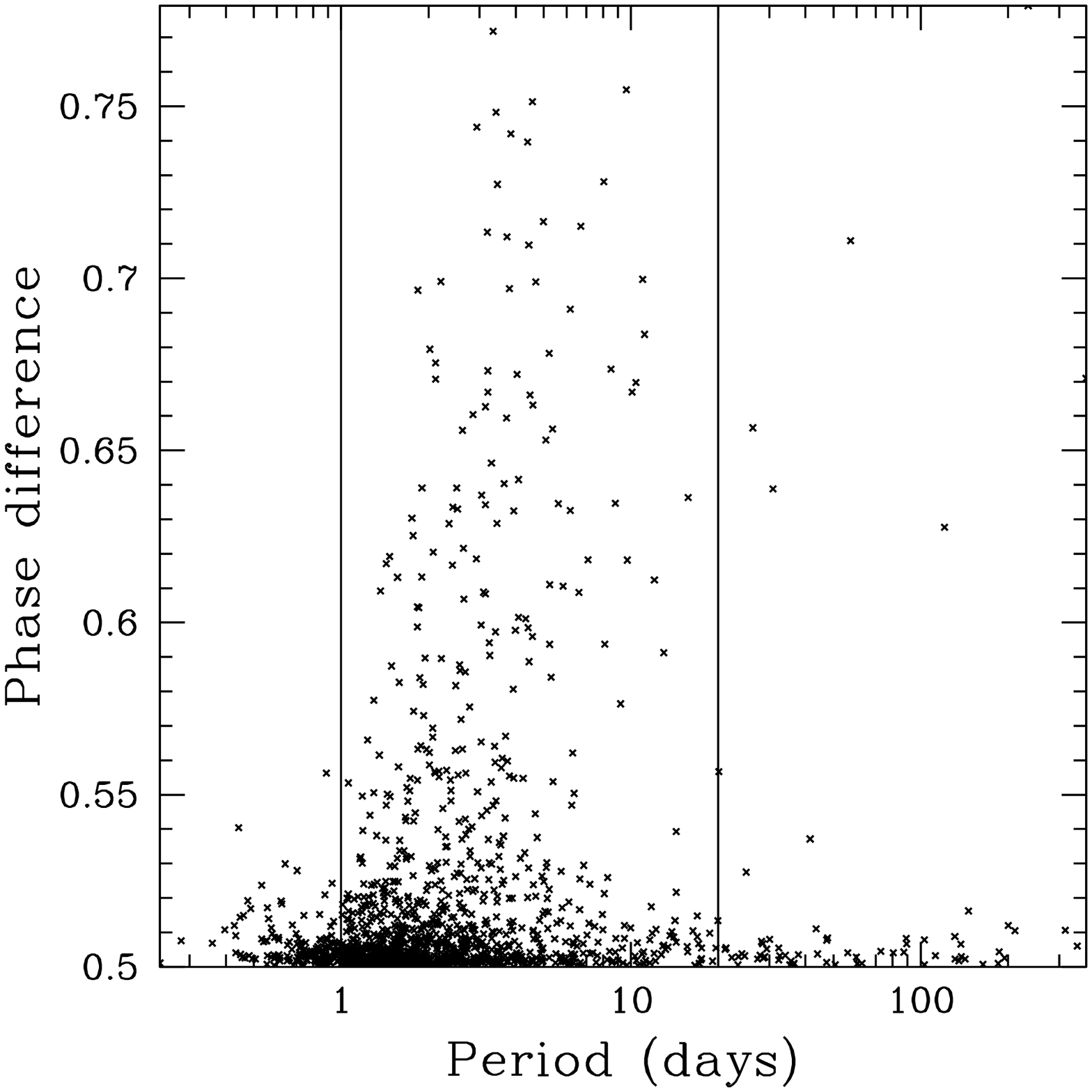}{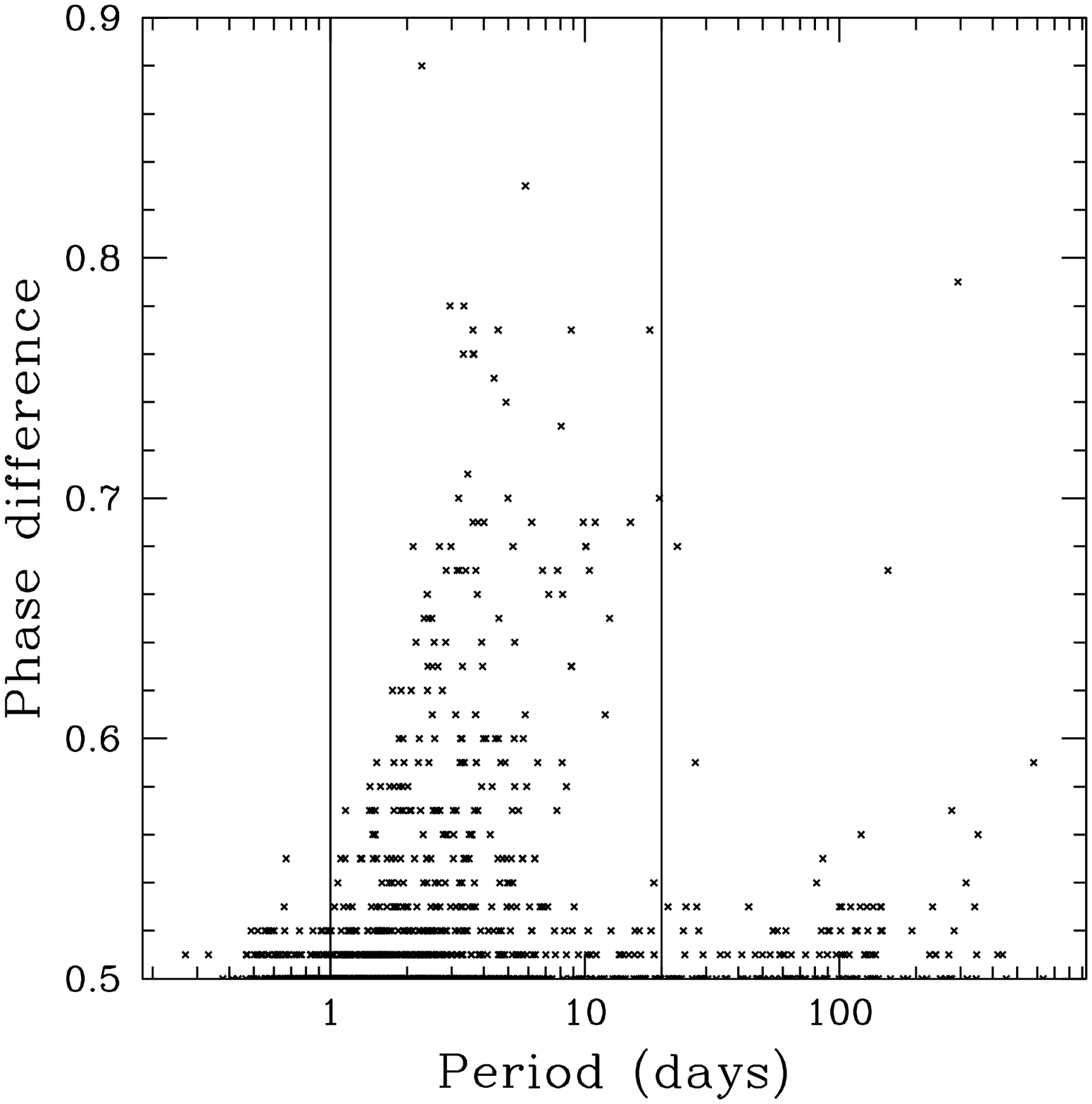}
\caption{
Period-Phase Difference Diagram for the four datasets considered.
\newline
Upper left: LMC MACHO sample ($4510$ EBs).
\newline
Upper right: LMC OGLE-II sample ($2474$ EBs).
\newline
Lower left: SMC MACHO sample ($1380$ EBs).
\newline
Lower right: SMC OGLE-II sample ($1317$ EBs).
\newline
For the LMC the circularization cutoff is evident at $P\sim 1.5\mathrm{d}$; for the SMC 
it is evident at $P\sim 1\mathrm{d}$.
}
\label{fig:perphd}
\end{center}
\end{figure}

The effects of period on eccentricity are further highlighted in Figure
\ref{fig:perphd} showing $|\phi_1-\phi_2|$ as a function of period for the four data sets we consider.
The LMC EBs from MACHO are shown in the upper left panel: the striking dependence of eccentricity on period for low periods is clearly shown and two ``steps'' at about $P\sim 0.5\mathrm{d}$ and $P\sim 1.5\mathrm{d}$ are clearly evident.
The lowest period objects in the LMC sample come mostly from contamination by foreground sources: there
are $\sim 60$ systems that, in view of their short periods ($P\lesssim 0.5\mathrm{d}$) and relatively 
high color ($\vr\sim 0.5~\mathrm{mag}$) are most likely composed of solar type stars located in the Milky Way \citep{faccioli07}. 
This is shown in Table \ref{tab:foreground} that reports data for the lowest period objects (defined as $P<0.5\mathrm{d}$) and shows that the most significant contribution to the 
first ``step'' at  $P\sim 0.5\mathrm{d}$ is largely due to this foreground population.
Figure \ref{fig:perphd} suggests that $P\sim 1.5 \mathrm{d}$ is the likely circularization cutoff period for the LMC and $P\sim 1 \mathrm{d}$ is the likely circularization cutoff period for the SMC; these cutoff periods are suggested both
both by the MACHO samples and the OGLE-II samples.
\tabletypesize{\footnotesize}
\begin{deluxetable}{lc}
\tablecolumns{6}
\tablewidth{0pc}
\tablecaption{Summary of very low period objects in the MACHO LMC sample.
\label{tab:foreground}}
\tablehead{
\colhead{Population} & 
\colhead{Number of objects}
}
\startdata
Very low period objects\tablenotemark{a} & $83$ \\

Very low period objects\tablenotemark{a} with circular orbit\tablenotemark{b} & $56$ \\

\hline
\hline

Foreground population & $62$ \\

Very low period objects\tablenotemark{a}~in the foreground population & $40$ \\

Very low period objects\tablenotemark{a} with circular orbit\tablenotemark{b}~in 
the foreground population & $36$ \\

\enddata
\tablenotetext{a}{Defined as $P< 0.5\mathrm{d}$.}
\tablenotetext{b}{Defined as $|\phi_1-\phi_2|< 0.51$.}
\end{deluxetable}
The absence of systems with eccentric orbits for $P\gtrsim 20\mathrm{d}$ is due to these systems being almost exclusively evolved and therefore having circularized their orbit on the Main Sequence.
The LMC EBs from OGLE-II are shown in the upper right panel of Figure \ref{fig:perphd} for the $2474$ EBs for which $|\phi_1-\phi_2|$ was provided; these are given to the second decimal place only and this accounts for the horizontal ``stripes'' in the OGLE-II panels in Figures \ref{fig:perphd} and \ref{fig:colphd} below.  The features are the same as the MACHO sample: we see again a ``step'' at 
$P\sim 1.5\mathrm{d}$ where there is a sharp increase in the range of eccentricities; EBs with eccentric orbits are almost absent for $P\gtrsim 20\mathrm{d}$; the almost complete absence of EB with $P\sim 0.5\mathrm{d}$ probably indicates an absence of foreground contamination in the OGLE-II sample.
The analogous result for the SMC are shown in the lower left panel of Figure \ref{fig:perphd} for the MACHO sample and in the lower right panel of Figure \ref{fig:perphd} for the OGLE-II sample.
The most notable feature of these diagrams is the different circularization cutoff: there
is a sharp increase in the range of eccentricities at $P\sim 1\mathrm{d}$ as opposed to
$P\sim 1.5\mathrm{d}$ for the LMC.
We explain this difference with the fact that EBs in the SMC are younger, on average,
than the LMC EBs; therefore only the shorter period systems, whose components are closer, have had enough time to circularize their orbit.
The fact that EBs in the two Clouds have, on average, populations at different stages of stellar evolution is shown in Table \ref{tab:smcvslmc}: the fraction of EBs belonging to the Main Sequence is considerably higher in the SMC than in the LMC.
There is also a population of $\sim 9$ EBs with 
$20\mathrm{d} \lesssim P\lesssim 100\mathrm{d}$ and $\vr\sim 0.3~\mathrm{mag}-0.5~\mathrm{mag}$ and strongly eccentric orbits.
The same qualitative features are shown by the OGLE-II sample; in particular the cutoff at
$P\sim 1\mathrm{d}$ is again clearly visible.
The OGLE-II SMC sample also reveal a sizeable ($\sim 20$) population of long period 
$P\gtrsim 80\mathrm{d}$, evolved ($1~\mathrm{mag}\lesssim V-I_{\mathrm{DIA}}\lesssim 2~\mathrm{mag}$) 
objects which are absent from the MACHO sample; several of these systems have
eccentric orbit; their light curves reveal that many of them are detached systems.
We attribute the fact that we did not find as many long period objects in the SMC MACHO sample to the fact that MACHO observed the SMC less frequently than the LMC, 
thus making less likely to find long period objects.
\normalsize
\par
The effects of color on eccentricity are shown in Figure \ref{fig:colphd} showing $|\phi_1-\phi_2|$ as a function of $\vr$ for the MACHO samples and of $V-I_{\mathrm{DIA}}$ for the OGLE-II samples.
The LMC EBs from MACHO are shown in the upper left panel; the Main Sequence corresponds there to $-0.2~\mathrm{mag}\lesssim\vr\lesssim 0.2~\mathrm{mag}$; the LMC EBs from OGLE-II are shown in the upper right panel for the $2474$ EBs for which $|\phi_1-\phi_2|$ was provided; in both cases it is evident that the bluest systems have the broadest range of eccentricities.
The SMC EBs from MACHO are shown in the lower left panel; the Main Sequence there corresponds again; to $-0.2~\mathrm{mag}\lesssim\vr\lesssim 0.2~\mathrm{mag}$ the LMC EBs from OGLE-II are shown in the lower right panel: in both cases is evident that the highest eccentricities are found in the bluest systems.
\begin{figure}
\footnotesize
\begin{center}
\plottwo{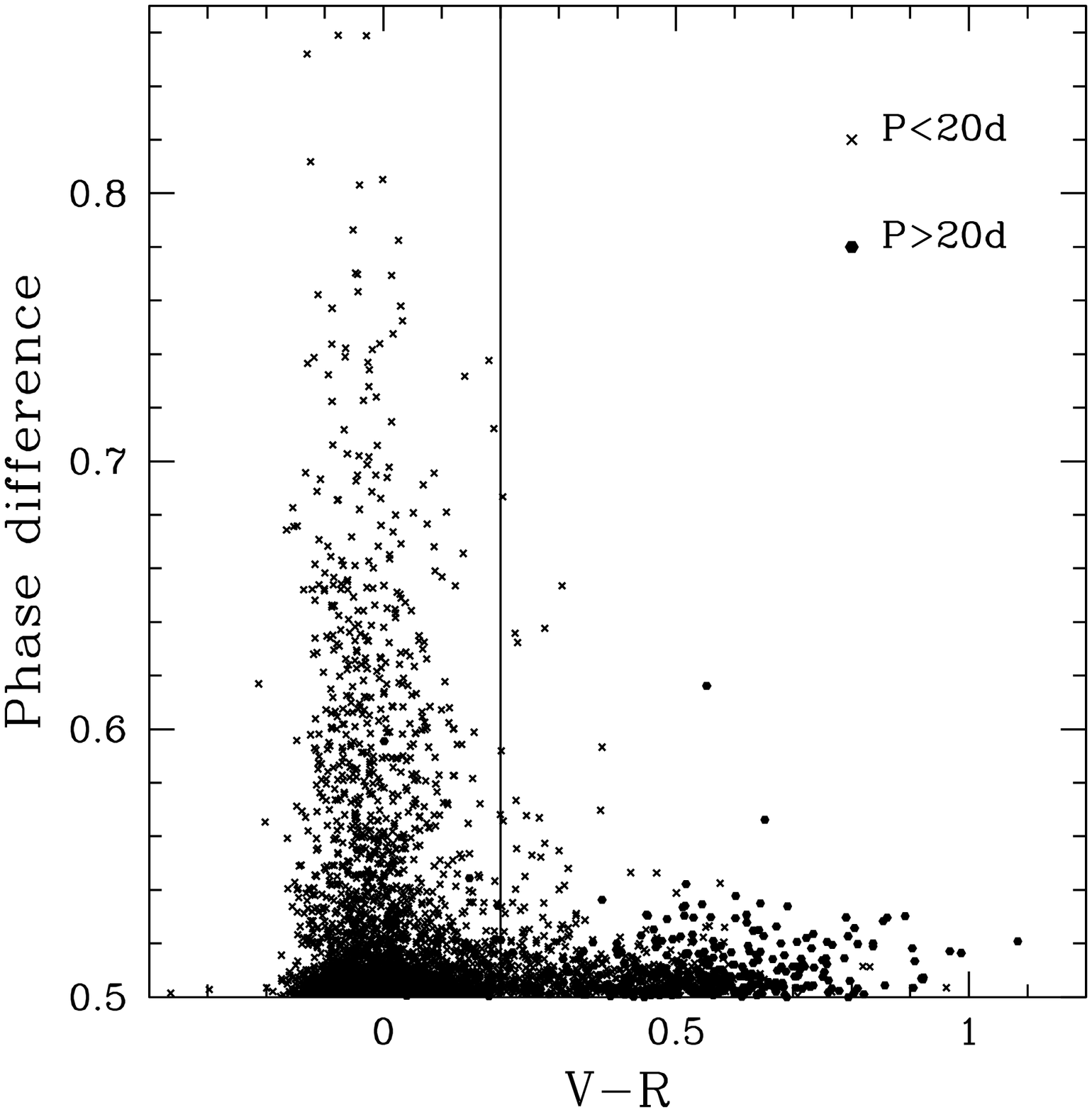}{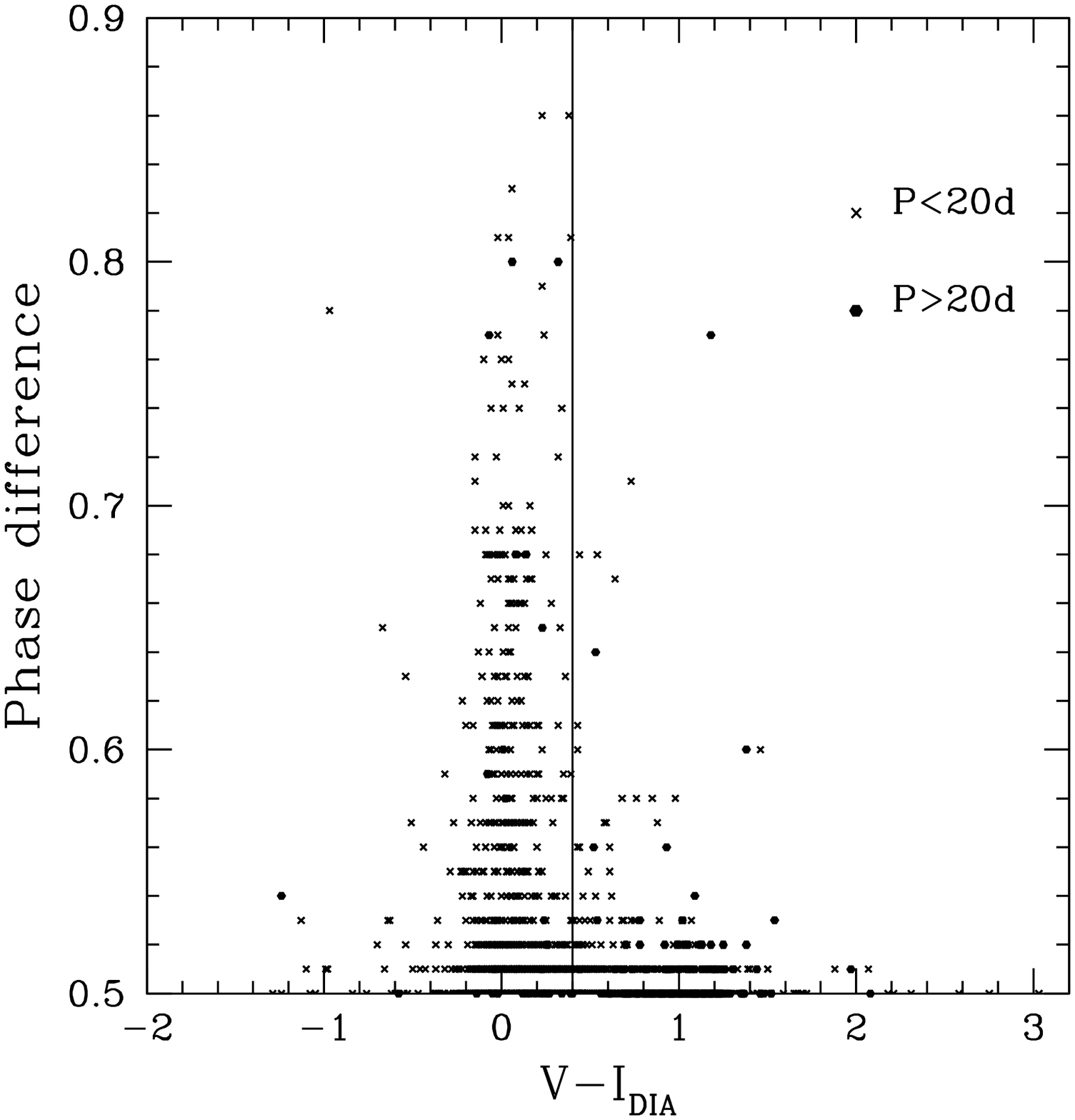}
\plottwo{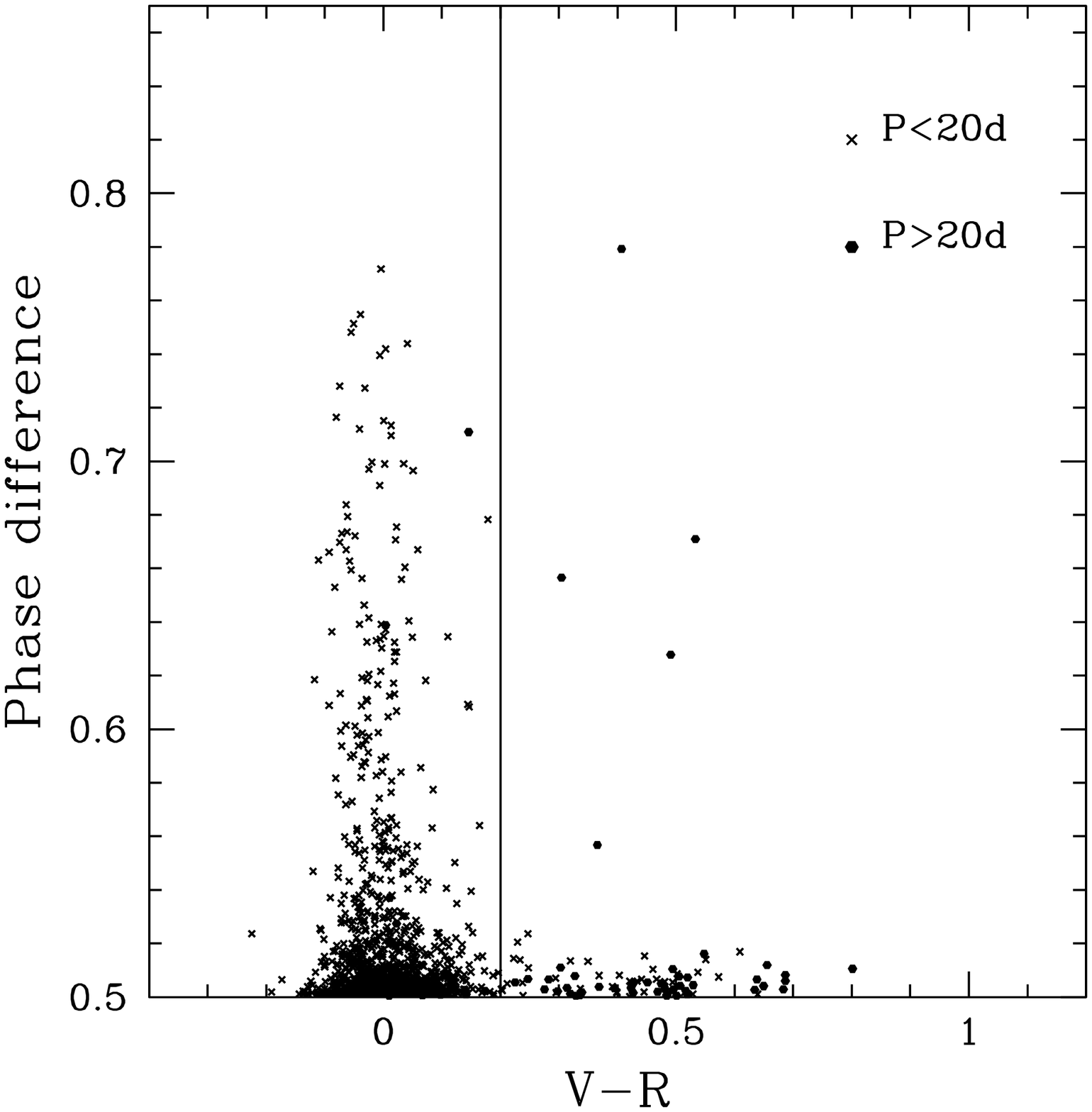}{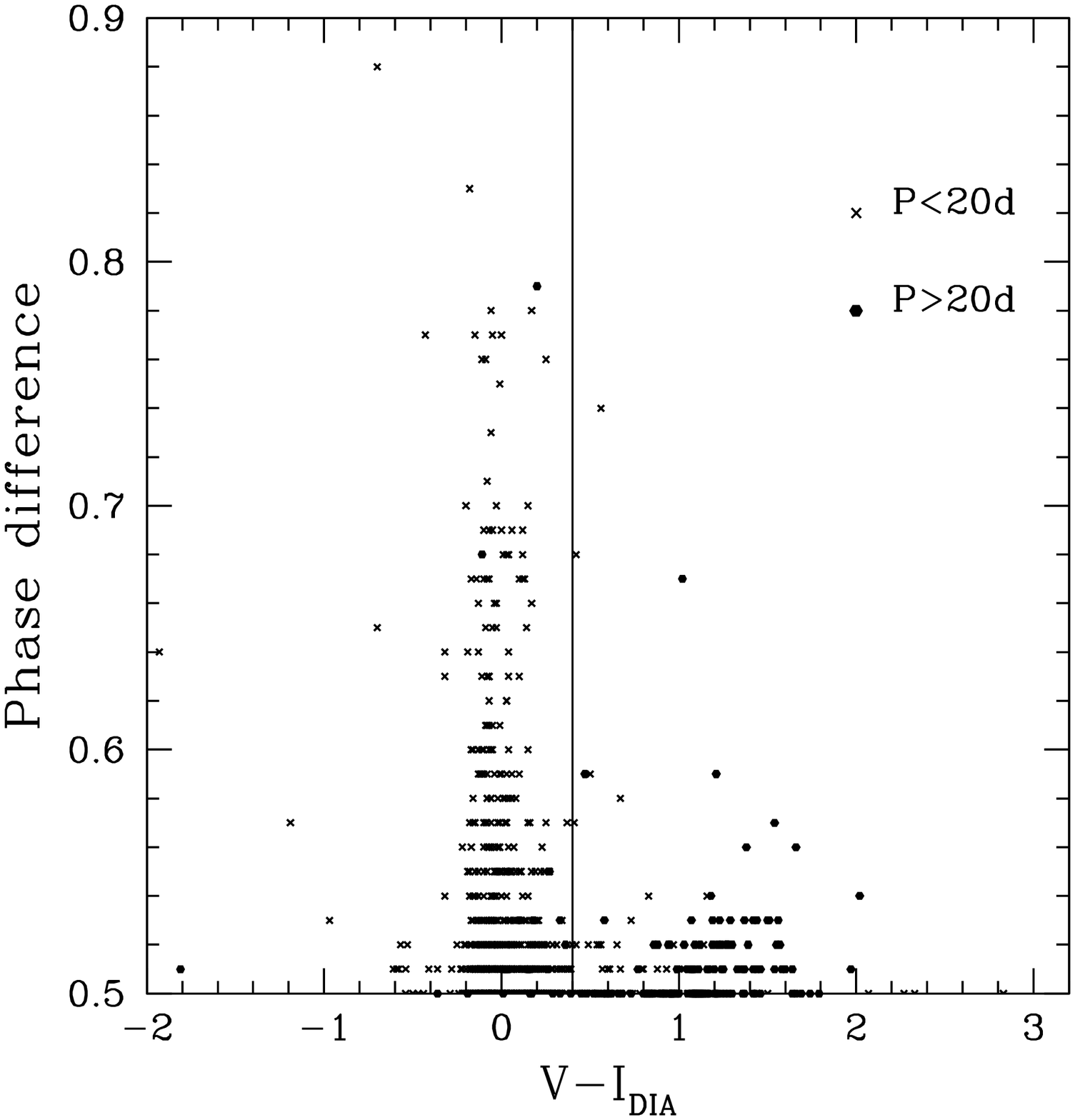}
\caption{ 
Color-Phase Difference Diagram for the four datasets considered.
\newline
Upper left: LMC MACHO sample ($4510$ EBs).
\newline
Upper right: LMC OGLE-II sample ($2474$ EBs). 
\newline
Lower left: SMC MACHO sample ($1380$ EBs). 
\newline
Lower right: SMC OGLE-II sample ($1317$ EBs).
}
\label{fig:colphd}
\end{center}
\end{figure}
\normalsize
\section{Discussion}
\label{sec:sig}
We assessed the significance of our results, in both Clouds, both for the MACHO and the OGLE-II samples.
In each case we proceeded in the following way.
First we selected two subsamples containing EBs with circular orbits and EBs with highly eccentric orbits respectively, and compared their colors via a Kolmogorov-Smirnov (KS) test \citep{press92}.
This test estimates the probability for two sets of random values to be drawn from the same distribution.
In our case we compare the distributions of colors, which we take as a rough proxy for evolutionary status since redder systems are generally more evolved.
This allows us to estimate the probability that, on average, the two subsamples contain systems in the same state of stellar evolution.
Figure \ref{fig:perphd} shows that the cutoff period above which elliptical orbits start to appear is $P\sim 1.5\mathrm{d}$ for the LMC and $P\sim 1\mathrm{d}$ for the SMC.
Furthemore above $P\sim20\mathrm{d}$ there are again almost only EBs with circular orbits in both Clouds.
Therefore we concentrated on the range $1.5\mathrm{d}<P<20\mathrm{d}$ for the LMC and
$1\mathrm{d}\le P \le 20\mathrm{d}$ for the SMC.
In these ranges we selected both EBs with circular orbits ($|\phi_1-\phi_2|<0.51$), and EBs with eccentric orbits ($|\phi_1-\phi_2|> 0.6\Rightarrow e>0.16$).
The cutoff of $0.51$ for circular orbits was chosen because, as Figures \ref{fig:errhist} and  \ref{fig:errhistrandomphase} show, the error in the phase difference is on average $0.01$ or less; thus a cutoff at $0.51$ usefully discriminates between circular and eccentric orbits; for the OGLE-II samples the cutoff is also appropriate since the phase of secondary minimum is reported up to two decimal places.
For the MACHO samples we used the $\vr$ color; for the OGLE-II samples we used both the $V-I$ color and to the $V-I_{\mathrm{DIA}}$ color.
\par
For the purpose of this discussion we call these subsamples the \emph{all ages subsamples} because they contain EBs of all ages; in particular they contain EBs that have evolved past the Main Sequence.
We then selected two more subsamples of \emph{blue} and \emph{young} EBs, again comprised of EBs with circular orbits and EBs with highly eccentric orbits.
These subsamples were selected with a color cut whose exact definition depends on the EB sample (see below), and with the same cuts in $|\phi_1-\phi_2|$ and $P$ as the all ages subsamples defined above and their colors were again compared via a KS test.
We call these subsamples the \emph{young EB subsamples} because the color cut ensures that they contain only young and unevolved stars that only recently settled on the Main Sequence.
Table \ref{tab:kstest} summarizes the properties of these subsamples; as the numbers show all subsamples are large enough for the KS test to be safely employed \citep{press92}; the
values of the KS statistic $D$ for all subsamples are reported in Table \ref{tab:kstest2}.
\par
\begin{figure}
\footnotesize
\begin{center}
\plottwo{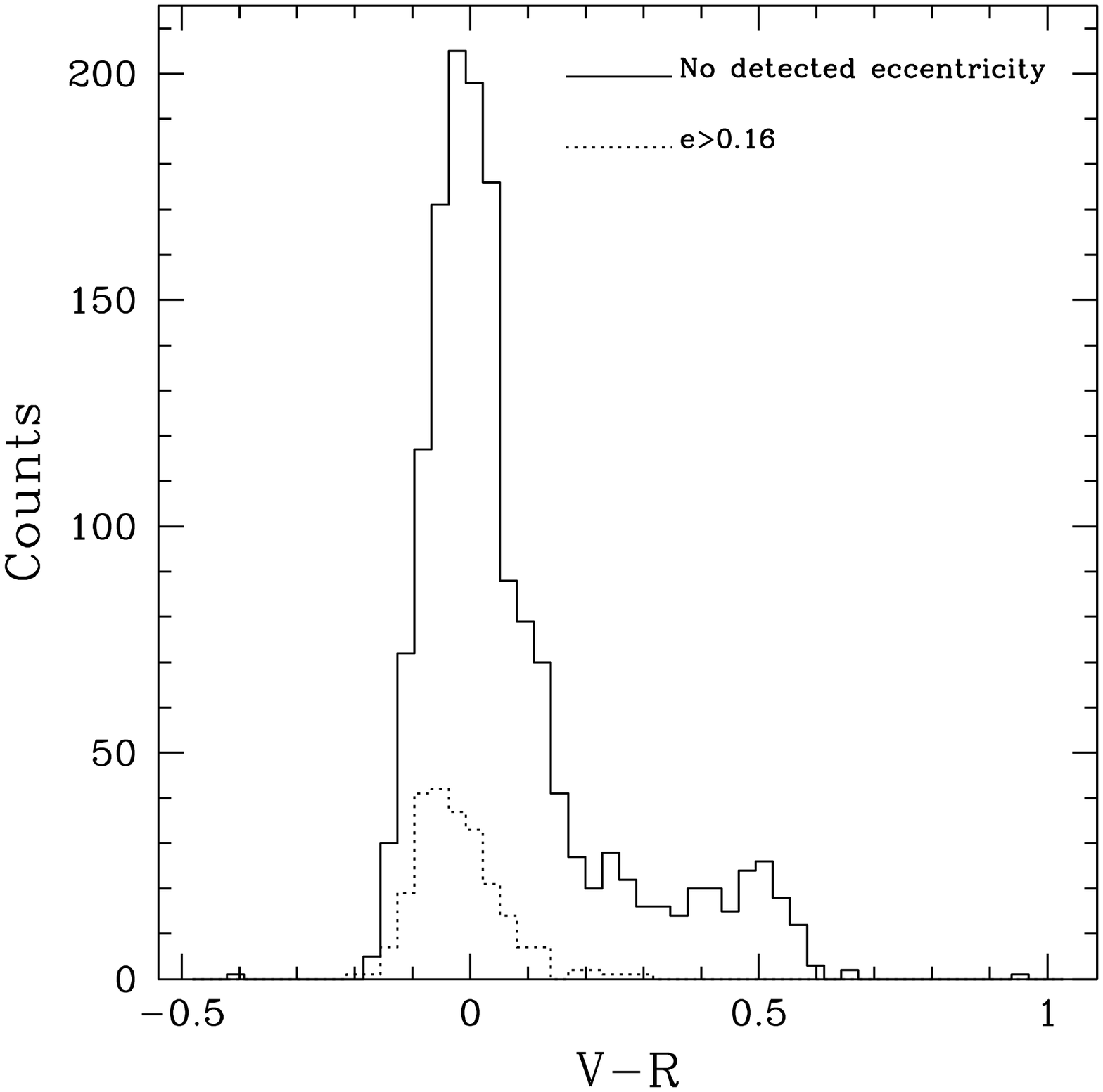}{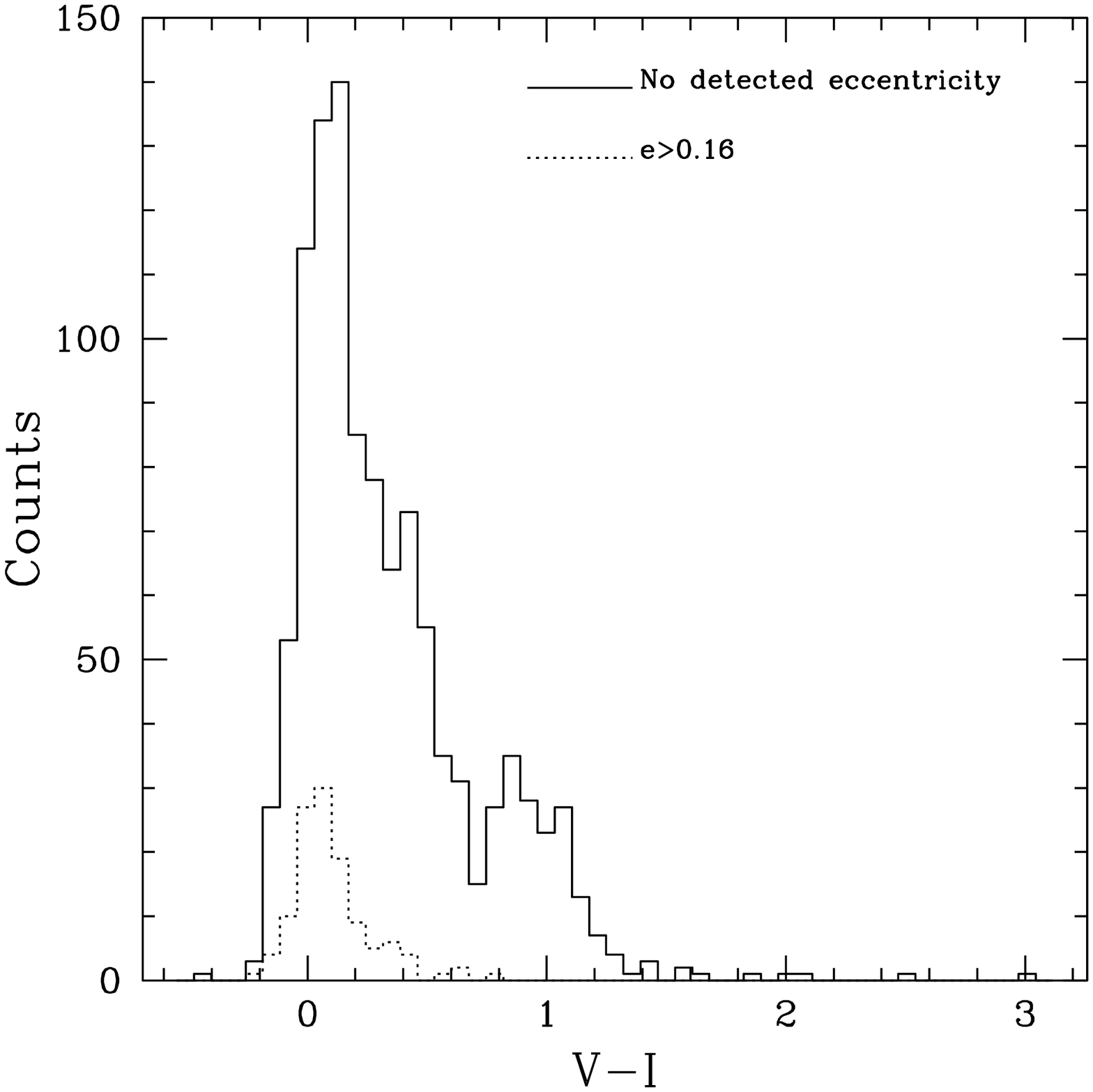}
\plottwo{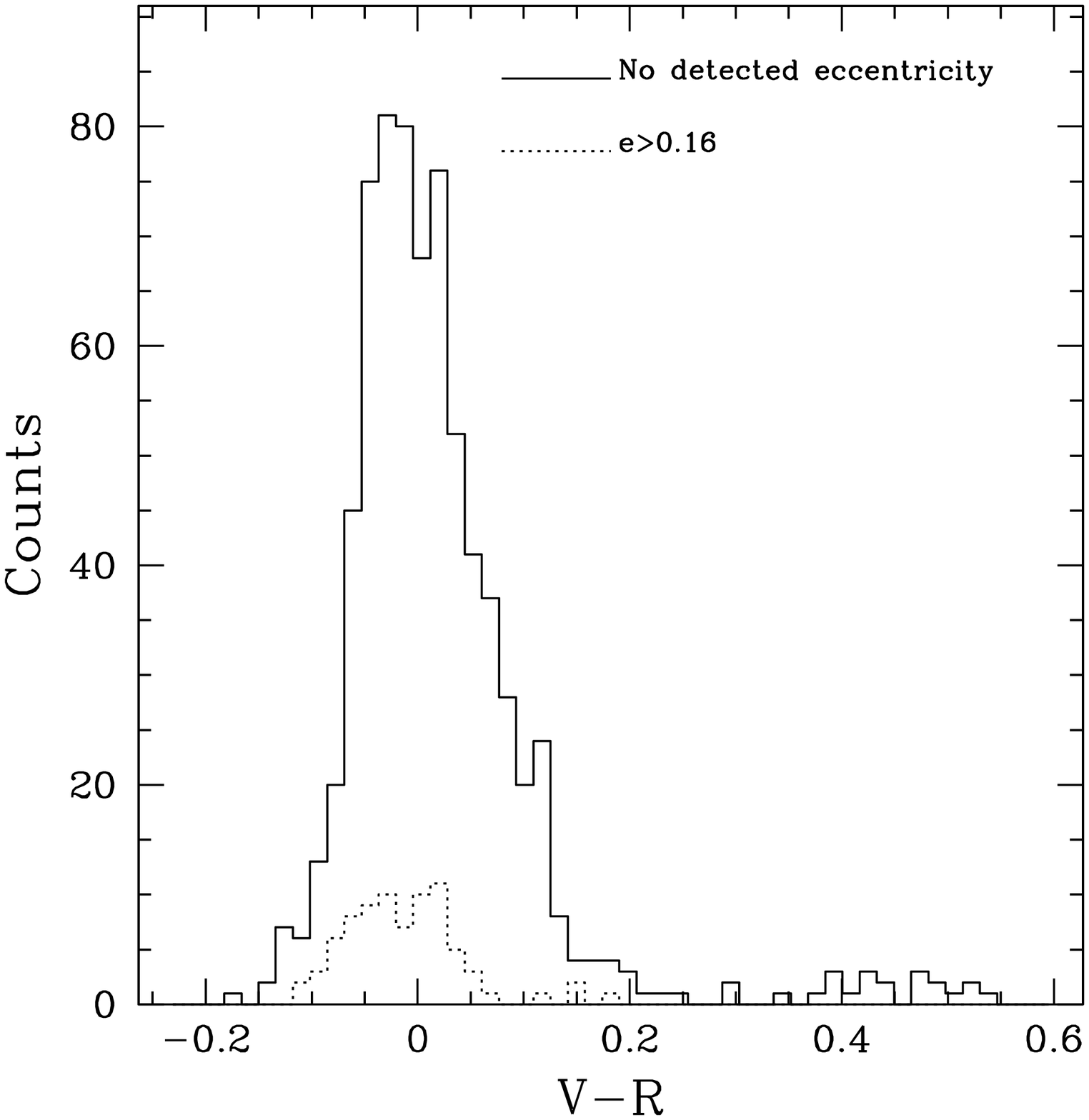}{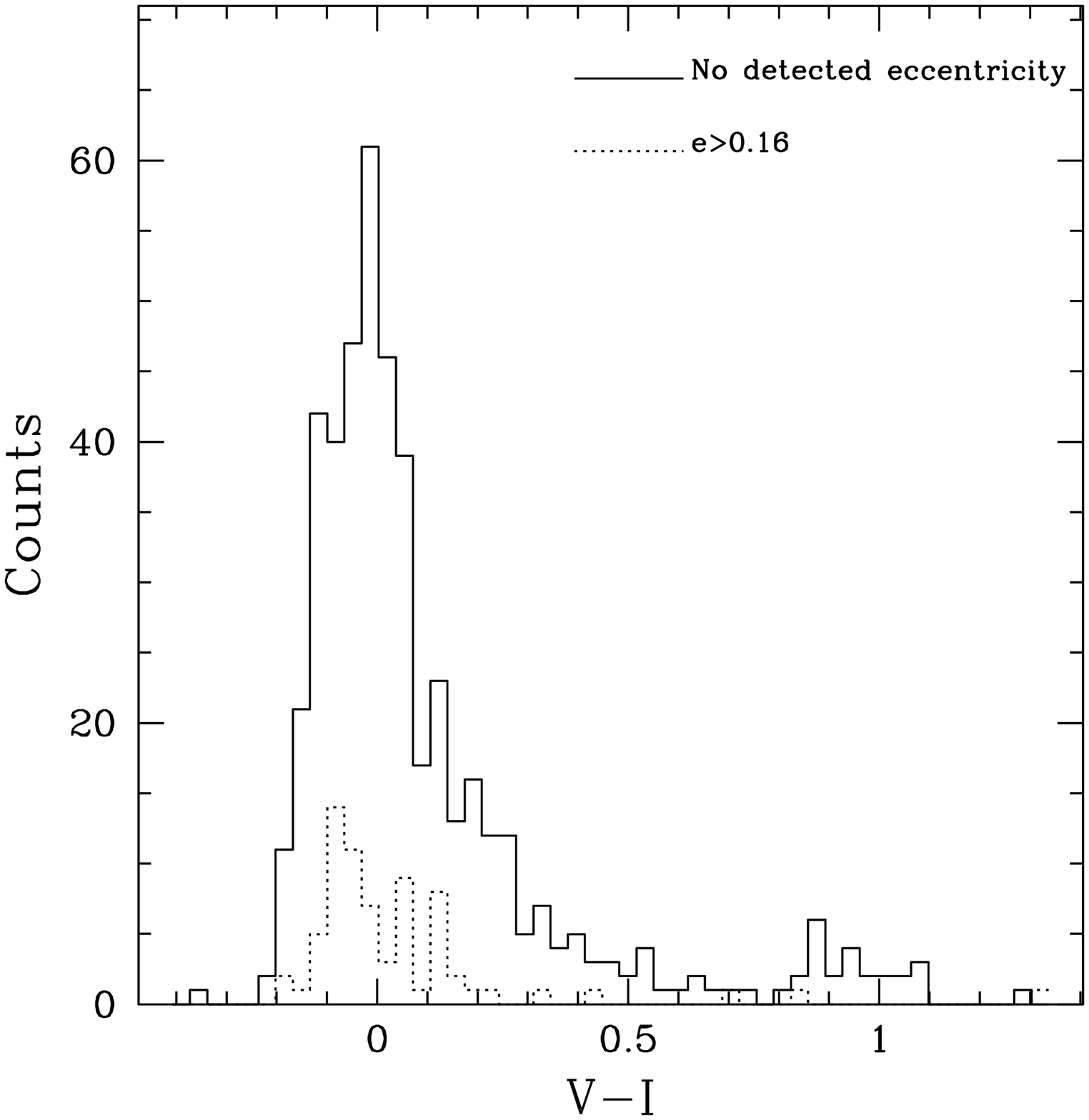}
\caption{
Color distribution for EBs with no detected eccentricity (continuous line) and for EBs with eccentricity $>0.16$ (dashed line) with $1.5\mathrm{d}<P<20\mathrm{d}$ for the LMC and
$1\mathrm{d}<P<20\mathrm{d}$ for the SMC.
Upper left: LMC EBs from MACHO.
Upper right: LMC EBs from OGLE-II.
Lower left: SMC EBs from MACHO.
Lower right: SMC EBs from OGLE-II.
}
\label{fig:coldist}
\end{center}
\end{figure}
\begin{figure}
\footnotesize
\begin{center}
\plottwo{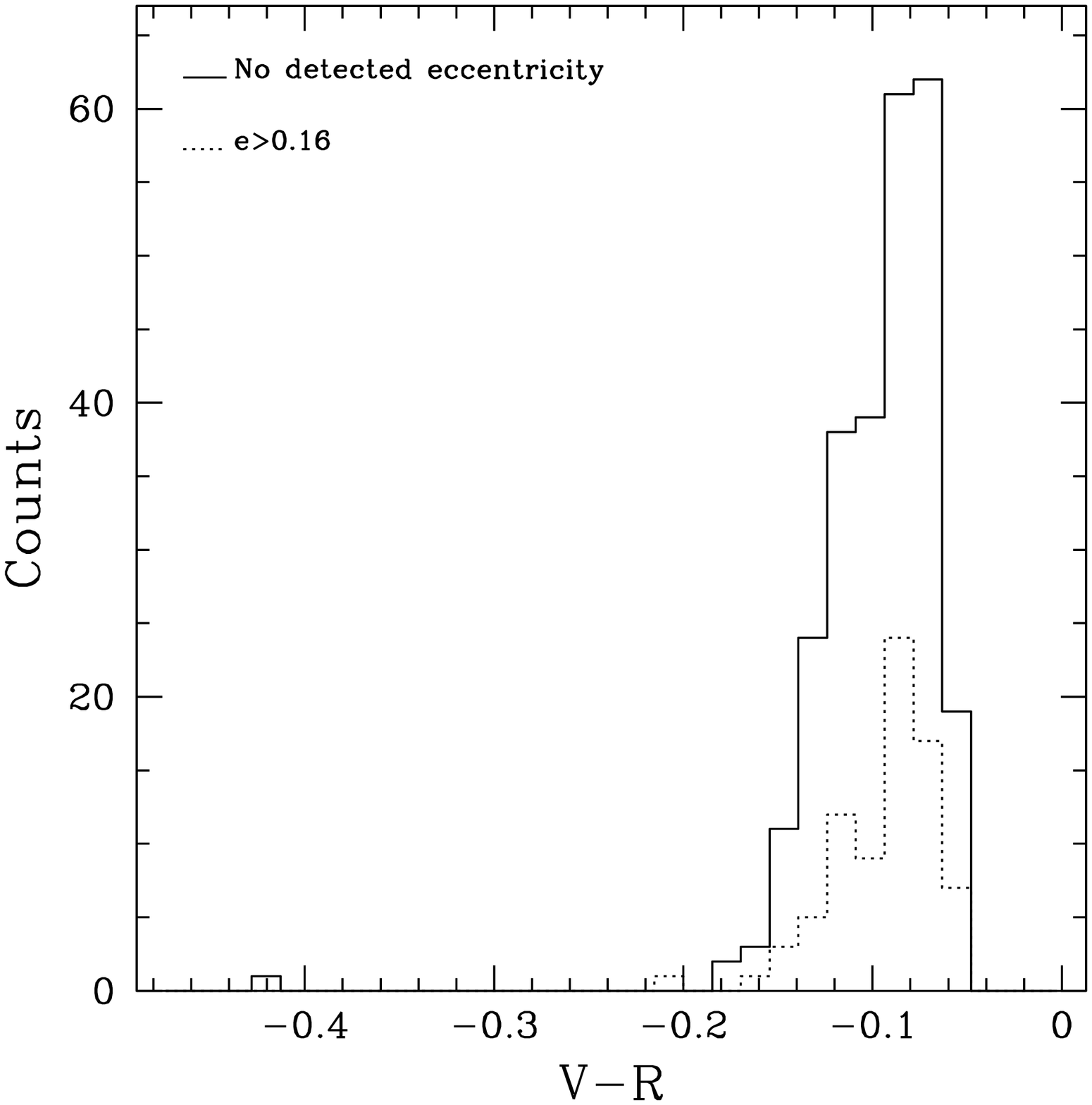}{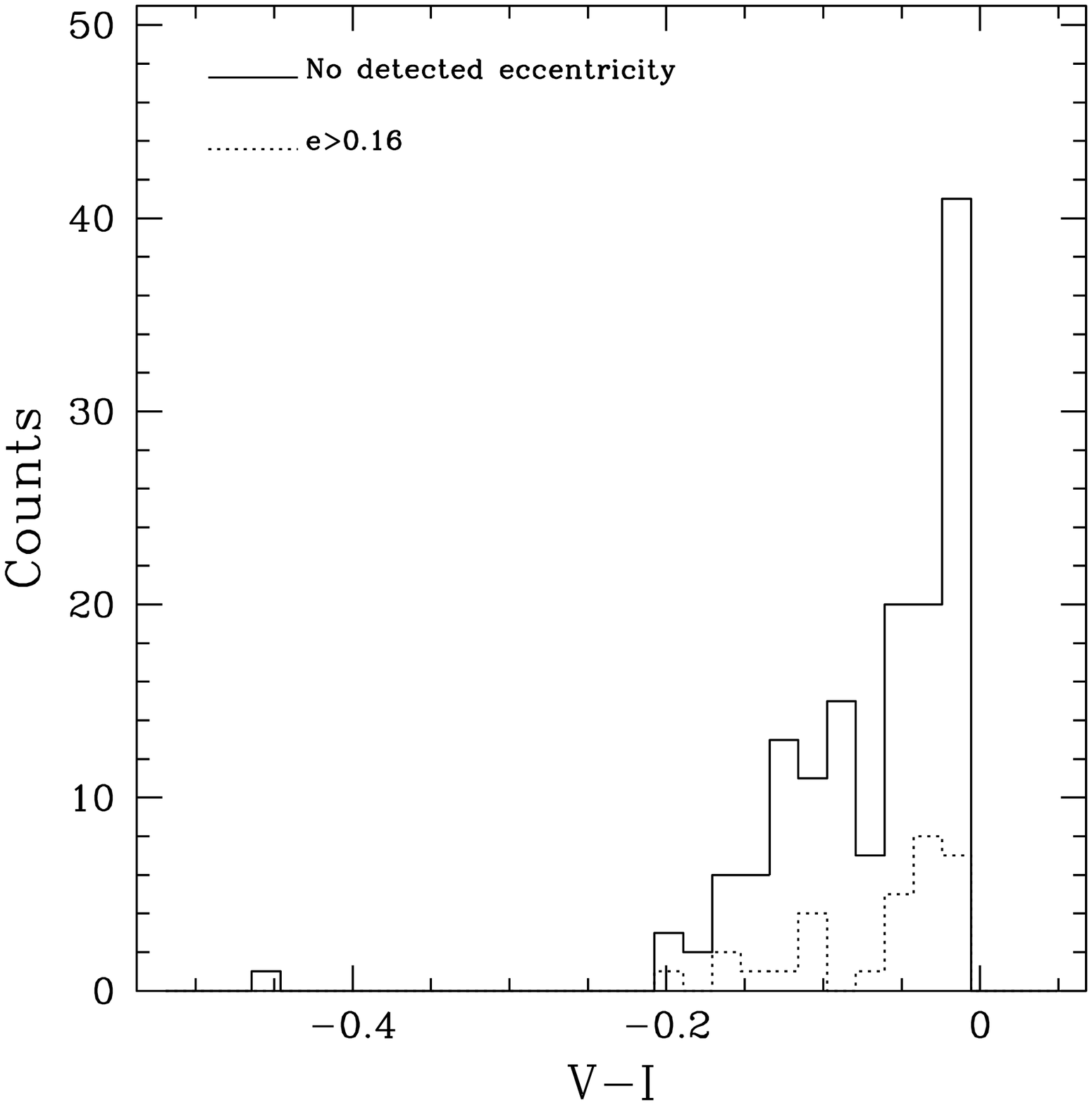}
\plottwo{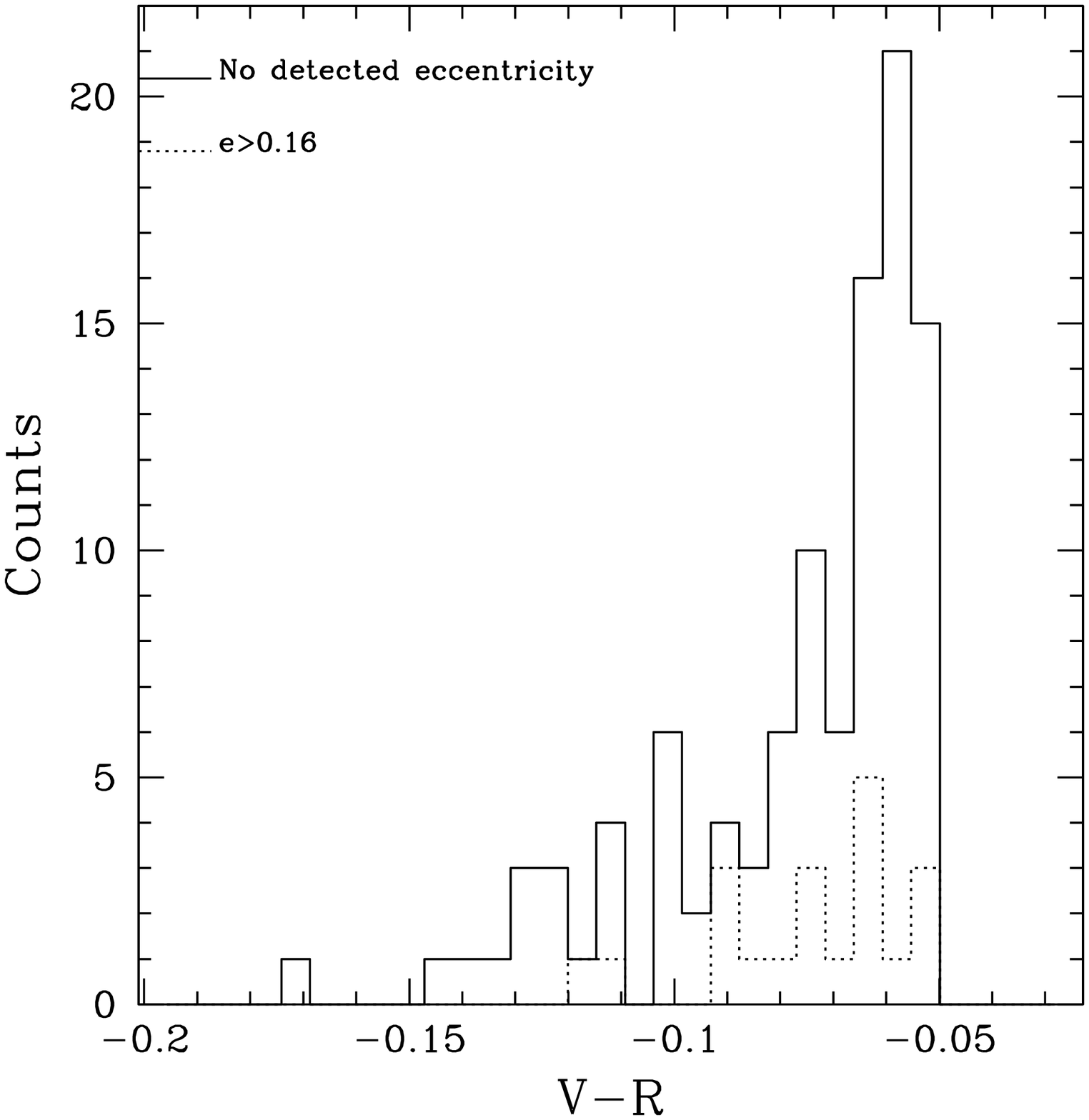}{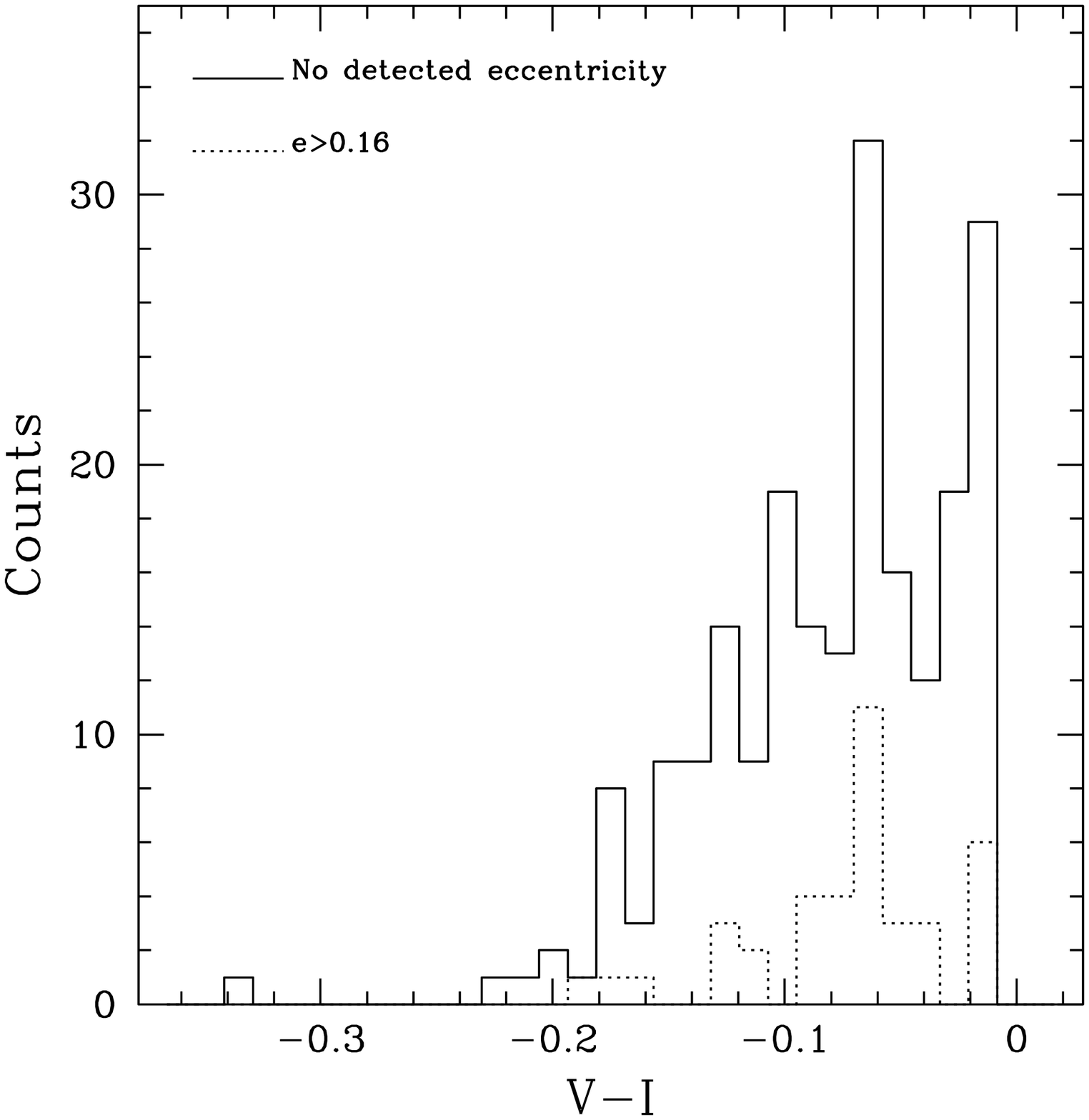}
\caption{
Color distribution for young EBs with no detected eccentricity (continuous line) and for EBs with eccentricity $>0.16$ (dashed line) with $1.5\mathrm{d}<P<20\mathrm{d}$ for the LMC and
$1\mathrm{d}<P<20\mathrm{d}$ for the SMC.
Upper left: LMC EBs from MACHO: $\vr<-0.06~\mathrm{mag}$.
Upper right: LMC EBs from OGLE-II: $V-I<0~\mathrm{mag}$.
Lower left: SMC EBs from MACHO: $\vr<-0.05~\mathrm{mag}$.
Lower right: SMC EBs from OGLE-II: $V-I<0~\mathrm{mag}$.
}
\label{fig:coldistcmdcutoff}
\end{center}
\end{figure}

The data reported in Table \ref{tab:kstest2} allow us to draw the main conclusion of this work, which hold for both Clouds, namely:
\begin{enumerate}
\item
There is a significant probability for the two young EB subsamples of being drawn from the same color distribution.
\item
The probability for the all ages subsamples of being drawn from the same distribution of colors is vanishingly small.
\end{enumerate}
These two findings allow us to state that, in both Clouds, EBs start their lives with a broad distribution of orbital eccentricities and circularize their orbits as their evolve on and past the Main Sequence.
This result is consistent with \citep{lucy71} who find that, in their sample of $103$ spectroscopic binaries, long period systems with late type giant components have mostly circular orbits; \citet{lucy71} suggest that tidal interactions due to expansion after the Main Sequence phase of one or both components is responsible for their circular orbit.
We note finally that the use of the $V-I$ cut in the OGLE-II samples gives results slightly more consistent with the results we obtain from the MACHO samples than the use of the $V-I_{\mathrm{DIA}}$ cut, although our conclusion does not qualitatively change.
For this reason in the histograms of Figures \ref{fig:coldist} and \ref{fig:coldistcmdcutoff} we show the $V-I$ color rather than the $V-I_{\mathrm{DIA}}$ color.
\tabletypesize{\footnotesize}
\begin{deluxetable}{ccccccc}
\tablecolumns{6}
\tablewidth{0pc}
\tablecaption{Summary of MACHO and OGLE-II EB eccentricity data.
\label{tab:kstest}}
\tablehead{
\colhead{Galaxy} & 
\colhead{Sample} &
\colhead{Circular\tablenotemark{a}} & 
\colhead{Eccentric\tablenotemark{b}} &
\colhead{Circular Young\tablenotemark{c}} &
\colhead{Eccentric Young\tablenotemark{c}} &
\colhead{Period Range}
}
\startdata
LMC & MACHO                              & $1537$ & $237$ & $260$ & $79$ & 
$1.5\mathrm{d}<P<20\mathrm{d}$ \\
LMC & OGLE-II ($I$ mags.)                & $1084$ & $119$ & $145$ & $30$ &
$1.5\mathrm{d}<P<20\mathrm{d}$ \\
LMC & OGLE-II ($I_{\mathrm{DIA}}$ mags.) & $1084$ & $119$ & $211$ & $44$ &
$1.5\mathrm{d}<P<20\mathrm{d}$ \\

\hline
\hline

SMC & MACHO                              & $724$ & $79$ & $104$ & $20$ &
$1\mathrm{d}<P<20\mathrm{d}$ \\
SMC & OGLE-II ($I$ mags.)                & $468$ & $70$ & $212$ & $39$ &
$1\mathrm{d}<P<20\mathrm{d}$ \\
SMC & OGLE-II ($I_{\mathrm{DIA}}$ mags.) & $468$ & $70$ & $233$ & $45$ &
$1\mathrm{d}<P<20\mathrm{d}$\\
\enddata
\tablenotetext{a}{Defined as $|\phi_1-\phi_2|<0.51$.}
\tablenotetext{b}{Defined as $|\phi_1-\phi_2|>0.6$.}
\tablenotetext{c}{Defined as $\vr<-0.06~\mathrm{mag}$ for the LMC MACHO sample, as $\vr<-0.05~\mathrm{mag}$ for the SMC MACHO sample, and as $V-I<0~\mathrm{mag}$ or $V-I_{\mathrm{DIA}}<0~\mathrm{mag}$ for both OGLE-II samples.}
\end{deluxetable}
\tabletypesize{\footnotesize}
\begin{deluxetable}{ccccc}
\tablecolumns{5}
\tablewidth{0pc}
\tablecaption{Results of KS tests for the subsamples of Table \ref{tab:kstest}.
\label{tab:kstest2}}
\tablehead{
\colhead{Galaxy} & 
\colhead{Sample} & 
\colhead{Subsample} & 
\colhead{KS Statistic $D$} &
\colhead{Probability of $D$}
}
\startdata
LMC & MACHO & All ages                               & $0.256$ & $2.3\times10^{-12}$ \\
LMC & OGLE-II & All ages ($I$ mags.)                 & $0.351$ & $3.3\times10^{-12}$ \\
LMC & OGLE-II & All ages ($I_{\mathrm{DIA}}$ mags.)  & $0.362$ & $5.7\times10^{-13}$ \\

    &         &                                      &         &        \\

LMC & MACHO & Young                                  & $0.092$ & $0.661$ \\
LMC & OGLE-II & Young ($I$ mags.)                    & $0.131$ & $0.757$ \\
LMC & OGLE-II & Young ($I_{\mathrm{DIA}}$ mags.)     & $0.157$ & $0.304$ \\

\hline
\hline

SMC & MACHO & All ages                               & $0.216$ & $0.002$ \\
SMC & OGLE-II & All ages ($I$ mags.)                 & $0.141$ & $0.162$ \\
SMC & OGLE-II & All ages ($I_{\mathrm{DIA}}$ mags.)  & $0.219$ & $0.005$ \\

    &         &                                      &         &         \\

SMC & MACHO & Young                                  & $0.156$ & $0.776$ \\
SMC & OGLE-II & Young ($I$ mags.)                    & $0.159$ & $0.354$ \\
SMC & OGLE-II & Young ($I_{\mathrm{DIA}}$ mags.)     & $0.186$ & $0.121$ \\

\enddata

\end{deluxetable}
\normalsize
\normalsize
\subsection{Influence of a third light}
One concern with our procedure is the possible influence of a third light.
If one EB system is close to a bright and blue undetected third star, the overall color 
of the system is skewed toward the blue.
This could result in an EB being erroneously selected as ``young'' by the color cuts described above, whereas the system is actually in a more advanced stage of evolution and biasing our
conclusions.
\par
We assessed the possible significance of this effect by fitting the $V$ light curves of the MACHO LMC EBs selected with the $\vr<-0.06~\mathrm{mag}$ cut, with with and without a third light as a free fit parameter; we then did the same for the SMC with the $\vr<-0.05~\mathrm{mag}$ cut.
If the $V$ band light curve fit improves considerably by using a third light as a free fit parameter, this means that an undetected blue third star is present; such system should therefore be excluded from the KS tests because the binary system is actually redder and probably more evolved than its $\vr$ suggests.
We point out that the only aim of these fits is to assess the significance of a possible undetected blue third star, and not to determine any physical parameters for the EBs.
\par
Since the systems we are dealing with are unevolved we expect them to be detached and therefore to be well described by the EBOP model \citep{etzel81,popper81}.
We actually performed the fits using the JKTEBOP\footnote{\url{http://www.astro.keele.ac.uk/\~{}jkt/codes/jktebop.html}}
\citep{southworth04a,southworth04b}
code, based on EBOP and which adds several modifications and extensions that make it easier to use, especially when fitting a large number of light curves.
We obtained preliminary starting values for the fit parameters using the 
DEBiL\footnote{\url{http://www.cfa.harvard.edu/\~{}jdevor/DEBiL.html}} code \citep{devor05}, a 
fitting code suited to automatically analyze large numbers of detached EBs.
We then refined the fits using JKTEBOP; this was done by performing a first fit using the parameters computed by DEBiL as starting values, then eliminating those points at least $3$ standard deviations away from this first fit and redoing the fit; this second step allowed in many cases to considerably improve the quality of the fit by excluding outlying points.
We used the following criterion to quantify whether or not the quality of the fit improved
as a result of fitting for a third light.
The typical number of points in the blue MACHO light curves is $\sim 690$ for the LMC and \
$\sim 840$ for the SMC and the number of fit parameters is $\sim 10$; we therefore assumed a typical number of degrees of freedom (dof) for our fits of $800$.
For $800$ dof, the probability of getting $\chi^2\ge 960\Rightarrow\chi^2/\mathrm{dof}\ge 1.2$ by 
chance is $<10^{-4}$ whereas the probability of getting 
$\chi^2\ge 800\Rightarrow\chi^2/\mathrm{dof}\ge 1$ by chance is $0.4933$.
Therefore, if for a $V$ light curve we have $\chi^2/\mathrm{dof}\ge 1.2$ without fitting for a third light and $\chi^2/\mathrm{dof}\le 1$ by fitting for it, we conclude that a blue third object is present; the EBs is then excluded by the young EB subsamples and the KS tests are performed again on them.
This procedure gave the subsamples described in Table \ref{tab:kstest3}; the results of the
KS tests are reported in Table \ref{tab:kstest4}.
\tabletypesize{\footnotesize}
\begin{deluxetable}{ccccccc}
\tablecolumns{6}
\tablewidth{0pc}
\tablecaption{Summary of MACHO eccentricity data for EBs with negligible influence of
a blue third light.
\label{tab:kstest3}}
\tablehead{
\colhead{Galaxy} & 
\colhead{Sample} &
\colhead{Circular\tablenotemark{a} Young\tablenotemark{c}} &
\colhead{Eccentric\tablenotemark{b} Young\tablenotemark{c}} &
\colhead{Period Range}
}
\startdata
LMC & MACHO & $240$ & $78$ & $1.5\mathrm{d}<P<20\mathrm{d}$ \\
SMC & MACHO & $99$ & $19$  & $1\mathrm{d}<P<20\mathrm{d}$ \\
\enddata
\tablenotetext{a}{Defined as $|\phi_1-\phi_2|<0.51$.}
\tablenotetext{b}{Defined as $|\phi_1-\phi_2|>0.6$.}
\tablenotetext{c}{Defined as $\vr<-0.06~\mathrm{mag}$ for the LMC MACHO sample, 
and as $\vr<-0.05~\mathrm{mag}$ for the SMC MACHO sample.}
\end{deluxetable}
\tabletypesize{\footnotesize}
\begin{deluxetable}{ccccc}
\tablecolumns{5}
\tablewidth{0pc}
\tablecaption{Results of KS tests for the subsamples of Table \ref{tab:kstest3}.
\label{tab:kstest4}}
\tablehead{
\colhead{Galaxy} & 
\colhead{Sample} & 
\colhead{Subsample} & 
\colhead{KS Statistic $D$} &
\colhead{Probability of $D$}
}
\startdata
LMC & MACHO & Young & $0.098$ & $0.597$ \\

SMC & MACHO & Young & $0.153$ & $0.816$ \\
\enddata
\end{deluxetable}
As Table \ref{tab:kstest4} shows, the probability of the two subsamples of being drawn from the same color distribution remain significant after accounting for the possible influence of a third light.
\subsection{Results from EROS}
\label{subsec:eros}
We considered the sample of $79$ EBs in the bar of the LMC found by the EROS collaboration \citep{gri95}.
We cross correlated this sample with both the MACHO sample, finding $42$ matches, and with
the OGLE-II sample finding $54$ matches; therefore this sample overlaps only partly with
the two larger samples and we performed the same analysis described above, finding comparable results, despite much lower number statistic.
Of the $79$ EBs in the sample we selected, in the $1.5\mathrm{d}<P<20\mathrm{d}$ period range, $33$ EBs with circular orbits and $9$ EBs with eccentric orbits.
We then applied the KS test to the distributions of the $B$ magnitudes for these two 
samples, since, despite their small size, the test can still be used \citep{press92}.
We found that the two distributions are different at $\sim 78\%$ level; the fact that the
confidence level is not higher can be attributed to small number statistic.
We did not attempt to select subsamples of young EBs since their numbers would have been too small for the KS test to be applied.
\section{Conclusion}
\label{sec:conclusion}
We have presented a study of orbital circularization in the Magellanic Clouds using EBs samples compiled both by the MACHO and by the OGLE-II collaboration, as well as a sample 
for the bar of the LMC compiled by the EROS collaboration.
We have shown that in the LMC binary stars with period in the range $1.5\mathrm{d}-20\mathrm{d}$ with initially eccentric orbits circularize during post Main Sequence evolution, with a cutoff in color at about $\vr<0.2~\mathrm{mag}$: our data are consistent with stars starting their lives with a wide range of eccentricities and circularizing their orbit as they evolve, both on the Main Sequence and in their red giant phase.
For the SMC the same conclusion holds, for a period range $1\mathrm{d}-20\mathrm{d}$.
We have obtained the same results with independently assembled datasets, thus enhancing our confidence in the validity of our conclusions.
\acknowledgments
This work uses public domain data from the MACHO project whose work was performed under the joint auspices of the U.S. Department of Energy, National Nuclear Security Administration by the University of California, Lawrence Livermore National Laboratory under contract No. W-7405-Eng-48, the National Science Foundation through the Center for Particle Astrophysics of the University of California under cooperative agreement AST-8809616, and the Mount Stromlo and Siding Spring Observatory, part of the Australian National University.
KHC's work is performed under the auspices of the U.S. Department of Energy
by Lawrence Livermore National Laboratory in part under Contract
W-7405-Eng-48 and in part under Contract DE-AC52-07NA27344.
This work uses public domain data obtained by the OGLE project.
LF acknowledges the kind hospitality of the Institute of Geophysics and Planetary Physics at Lawrence Livermore National Laboratory and of the Harvard-Smithsonian Center for Astrophysics where part of the work was done.
We also thank the referee for helpful suggestions, and we thank Leon Lucy for pointing out to us the reference \citep{lucy71}.
\newpage
\end{document}